\documentclass{article}
\usepackage{xcolor}         
\usepackage[preprint]{style}

\usepackage[utf8]{inputenc} 
\usepackage[T1]{fontenc}    
\usepackage{hyperref}       
\usepackage{url}            
\usepackage{booktabs}       
\usepackage{amsfonts}       
\usepackage{nicefrac}       
\usepackage{microtype}      
\usepackage{enumitem}
\usepackage{times}
\usepackage{latexsym}
\usepackage{graphicx}
\usepackage{multicol}
\usepackage{multirow}
\usepackage{amsmath}
\usepackage{amssymb}
\usepackage{bbding}
\usepackage{subfigure}
\usepackage{colortbl}
\usepackage{pifont}
\usepackage{makecell}
\usepackage{mathrsfs}
\usepackage{bm}
\usepackage{bbm}
\usepackage{dsfont}
\usepackage{pgf}
\usepackage{fontawesome}
\usepackage{wrapfig}
\usepackage{tcolorbox}

\usepackage{caption}

\definecolor{darkgray}{gray}{0.3}
\definecolor{lightblue}{RGB}{115, 192, 222}

\newtcolorbox{promptbox}[2][Prompt]{
colback=black!5!white,
arc=4pt, 
boxrule=0.5pt,
fonttitle=\bfseries,
title=#1, 
before upper={\small}, fontupper=\fontfamily{ptm}\selectfont,
colframe=#2,
}

\def\colornum#1{
\newcommand{\MidNumber}{38}  
\newcommand{\MaxNumber}{100} 
\newcommand{\MinNumber}{0}   
\pgfmathsetmacro{\PercentColor}{max(min(100.0*(#1 - \MinNumber)/(\MaxNumber-\MinNumber),100.0),0.00)}
\hspace{-0.33em}\colorbox{pink!\PercentColor!white}{#1}
}

\newcommand{\ourmethod}{\textsc{mCoder}}

\newcommand{\benchmark}{\textsc{McEval}}
\newcommand{\instruct}{\textsc{McEval-Instruct}}

\title{\benchmark{}: Massively Multilingual Code Evaluation}
\author{
  Linzheng Chai\textsuperscript{\rm 1 \thanks{\ \ Equal contribution. }}, 
  Shukai Liu\textsuperscript{\rm 1 *},
  Jian Yang\textsuperscript{\rm 1 *}\thanks{\ \ Corresponding Author.}, 
  Yuwei Yin\textsuperscript{\rm 2}, 
  {\bf Ke Jin}\textsuperscript{\rm 1}, 
  {\bf Jiaheng Liu}\textsuperscript{\rm 1},
  \\{\bf Tao Sun}\textsuperscript{\rm 1}, 
  {\bf Ge Zhang}\textsuperscript{\rm 3},
  {\bf Changyu Ren}\textsuperscript{\rm 1},  
  {\bf Hongcheng Guo}\textsuperscript{\rm 1}, 
  {\bf Zekun Wang}\textsuperscript{\rm 1}, 
  {\bf Boyang Wang}\textsuperscript{\rm 1},
  \\
  {\bf Xianjie Wu}\textsuperscript{\rm 1}, 
  {\bf Bing Wang}\textsuperscript{\rm 1},
  {\bf Tongliang Li}\textsuperscript{\rm 4}, 
  {\bf Liqun Yang}\textsuperscript{\rm 1}, 
  {\bf Sufeng Duan}\textsuperscript{\rm 5}, 
  {\bf Zhoujun Li}\textsuperscript{\rm 1} \\
 \textsuperscript{\rm 1}CCSE, Beihang University,~\textsuperscript{\rm 2}University of British Columbia,~\textsuperscript{\rm 3}University of Waterloo\\
 \textsuperscript{\rm 4}Beijing Information Science and Technology University,~\textsuperscript{\rm 5}Shanghai Jiao Tong University\\
  \texttt{challenging@buaa.edu.cn} \\
}

\begin{document}

\maketitle
\vspace{-10pt}
\begin{abstract}
Code large language models (LLMs) have shown remarkable advances in code understanding, completion, and generation tasks. Programming benchmarks, comprised of a selection of code challenges and corresponding test cases, serve as a standard to evaluate the capability of different LLMs in such tasks.
However, most existing benchmarks primarily focus on Python and are still restricted to a limited number of languages, where other languages are translated from the Python samples (e.g. MultiPL-E) degrading the data diversity.
To further facilitate the research of code LLMs, we propose a massively multilingual code benchmark covering 40 programming languages (\benchmark{}) with 16K test samples, which substantially pushes the limits of code LLMs in multilingual scenarios.
The benchmark contains challenging code completion, understanding, and generation evaluation tasks with finely curated massively multilingual instruction corpora \instruct{}. In addition, we introduce an effective multilingual coder \ourmethod{} trained on \instruct{} to support multilingual programming language generation.
Extensive experimental results on \benchmark{} show that there is still a difficult journey between open-source models and closed-source LLMs (e.g. GPT-series models) in numerous languages. The instruction corpora, evaluation benchmark, and leaderboard are available at \url{https://mceval.github.io/}.
\end{abstract}

\vspace{-15pt}
\section{Introduction}
\label{sec:introduction}
\vspace{-10pt}
Large language models (LLMs) designed for code, such as Codex~\citep{codex}, CodeGen~\citep{codegen}, Code Llama~\citep{code_llama}, DeepSeekCoder~\citep{deepseek_coder}, and CodeQwen~\citep{Qwen} excel at code understanding, completion, and generation tasks.
Code LLMs with a large number of parameters (e.g. 7B, 13B, or larger) are pre-trained on large-scale code databases with self-supervised autoregressive objectives, followed by instruction tuning~\citep{instruct_gpt} for aligning to human preferences and downstream code-related tasks.


Most code benchmarks~\citep{codex,mbpp,mbxp} are introduced to evaluate the performance of code LLMs by assessing their ability to generate executable code based on the problem descriptions. The assessments aim to gauge the capacity of the models to understand and generate code effectively, thereby contributing to facilitating and streamlining the programming process for developers. The execution-based method executes generated code against test cases to measure the success rate. Due to the difficulty of creating the problem and its corresponding solution (requiring specialized programming staff), the development of evaluation benchmarks is limited within Python, with a few other languages being translated from Python.
\textit{Therefore, the community desperately needs a massively multilingual programming benchmark (not from HumanEval or MBPP) comprised of instruction corpora and evaluation set to comprehensively facilitate and evaluate the generation, completion, and understanding capability of LLMs.}

\begin{wrapfigure}{r}{0.5\textwidth}
    \centering
    \includegraphics[width=0.5\textwidth]{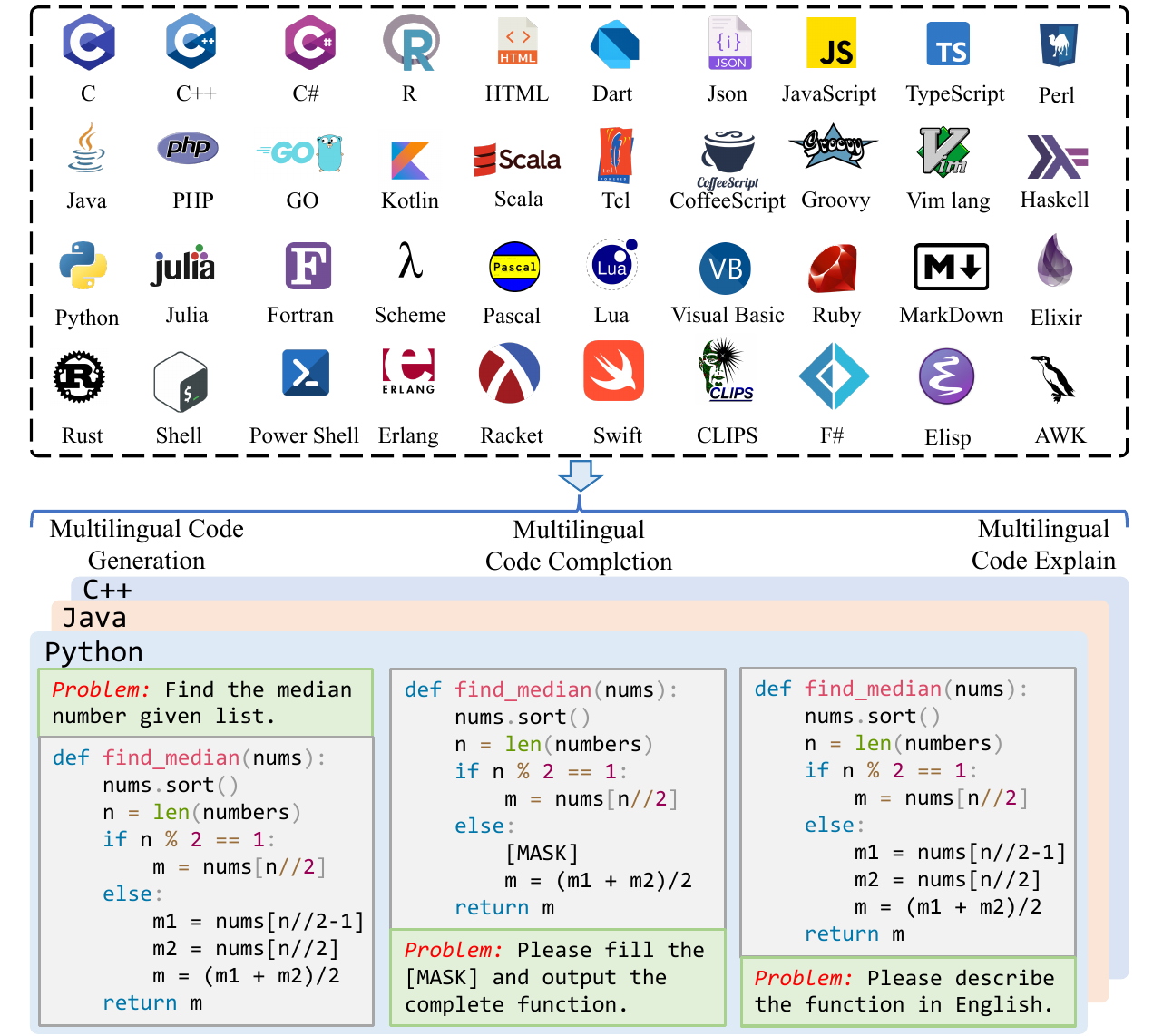}
    \vspace{-5mm}
    \caption{Massively multilingual evaluation task comprised of three tasks, including code generation, code completion, and code explanation.}
    \label{fig:intro}
    \vspace{-4mm}
\end{wrapfigure}

To facilitate the development of code LLMs, we introduce a complete framework that includes the multilingual code instruction corpora, multilingual coder (\ourmethod{}), and multilingual code evaluation benchmark. 
First, we propose \benchmark{}, the first massively multilingual code evaluation benchmark (from human handwriting) covering 40 languages (16K samples in total), encompassing multilingual code generation, multilingual code explanation, and multilingual code completion tasks. Then, we create a massively multilingual instruction corpora \instruct{} of 40 languages. We initially select and refine high-quality code snippets from various programming languages (PLs) using an LLM. The LLM then generates clear and self-contained instructional content, including problem descriptions and corresponding solutions, based on the refined snippets. To ensure consistency and enhance learning across languages, we introduce cross-lingual code transfer, adapting instructional content to different PLs while increasing sample complexity. Based on open-source models and \instruct{}, \ourmethod{}  is used as a strong baseline to explore the transferability of LLMs among different PLs.

The contributions are summarized as follows: (1) We propose \benchmark{} with enough test samples (16K), a true massively multilingual multitask code evaluation benchmark (not from HumanEval or MBPP) covering 40 languages, encompassing multilingual code generation, multilingual code explanation, and multilingual code completion tasks. 
(2) We introduce \instruct{}, the massively multilingual code instruction corpora covering the multilingual code snippet from 40 languages. Based on \instruct{}, an effective multilingual coder \ourmethod{} is used as a strong baseline for \benchmark{}.
(3) We systematically evaluate the understanding and generation capabilities of 20+ models on our created \benchmark{} and create a leaderboard to evaluate them on 40 programming languages dynamically. Notably, extensive experiments suggest that comprehensive multilingual multitask evaluation can realistically measure the gap between open-source (e.g. DeepSeekCoder and CodeQwen1.5) and closed-source models (e.g. GPT-3.5 and GPT-4).

\vspace{-10pt}

\section{Multilingual Code Evaluation: \benchmark{}}

\vspace{-8pt}

\subsection{Dataset Statistics}
\begin{figure*}[t]
\begin{center}
    \includegraphics[width=0.85\textwidth]{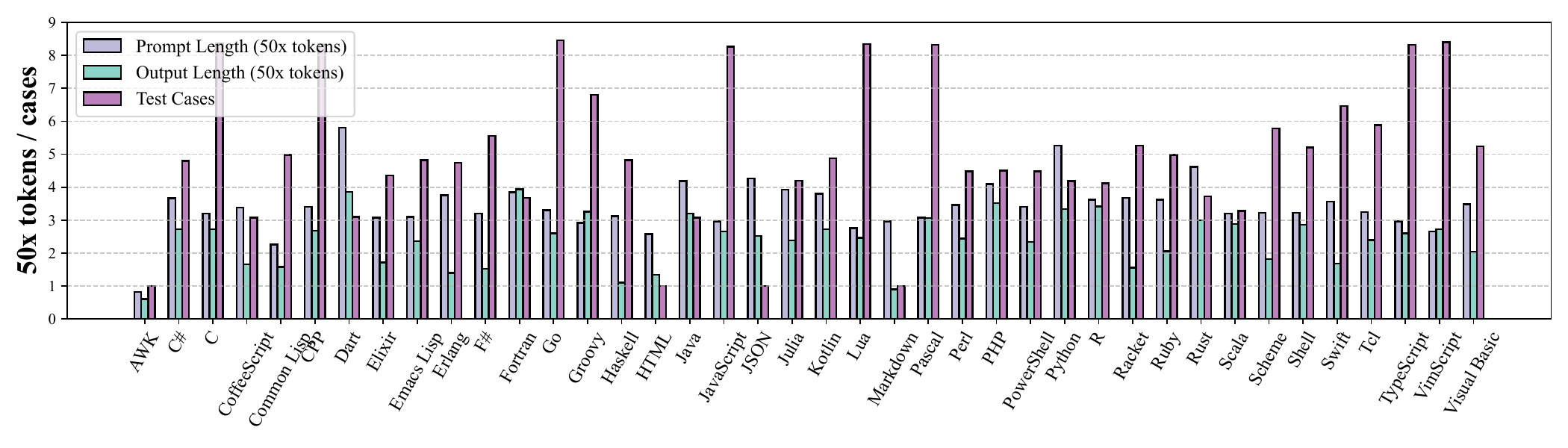}
    \vspace{-5pt}
    \caption{Data statistics of the proposed \benchmark{} benchmark.}
    \label{fig:data_statistics}
    \vspace{-10pt}
\end{center}
\end{figure*}

The created \benchmark{} is comprised of three key code-related tasks covering 40 programming languages, including multilingual code generation, multilingual code explanation, and multilingual code completion tasks. \autoref{fig:data_statistics} plots the length of input length, the length of output, and the number of test cases of each programming language.
The multilingual code generation and explanation tasks separately contain 2K samples, where each language has nearly 50 samples.
The code completion task can be decomposed into\textit{multi-line completion} (3K samples), \textit{single-line completion} (3K samples), \textit{span completion} (4K samples), and \textit{span completion (light)} (2K samples)~\citep{fim}.

\vspace{-8pt}
\subsection{Human Annotation \& Quality Control}
To create the massively multilingual code evaluation benchmark \benchmark{}, the annotation of multilingual code samples is conducted utilizing a comprehensive and systematic human annotation procedure, underpinned by rigorously defined guidelines to ensure accuracy and consistency. Initially, 10 software developers in computer science are recruited as multilingual programming annotators with proven proficiency in the respective programming languages. Following a detailed training session on the annotation protocol, which emphasizes the importance of context, syntactical correctness, and semantic fidelity across languages, annotators are tasked with creating problem definitions and the corresponding solution. The annotators should follow: (1) Provide a clear and self-contained problem definition, answer the question with any tools, and design the test cases to evaluate the correctness of the code. (2) Classify them into multiple difficulties (Easy/Middle/Hard), based on algorithmic complexity and functionality.
Each sample is independently annotated by at least two annotators to minimize subjective bias and errors. Discrepancies between annotators are resolved through consensus or adjudication by a senior annotator. Finally, three volunteers are employed to evaluate the correctness of the benchmark (> 90\% accuracy) and correct the errors.
\begin{figure*}[t!]
\begin{center}
    \includegraphics[width=0.95\textwidth]{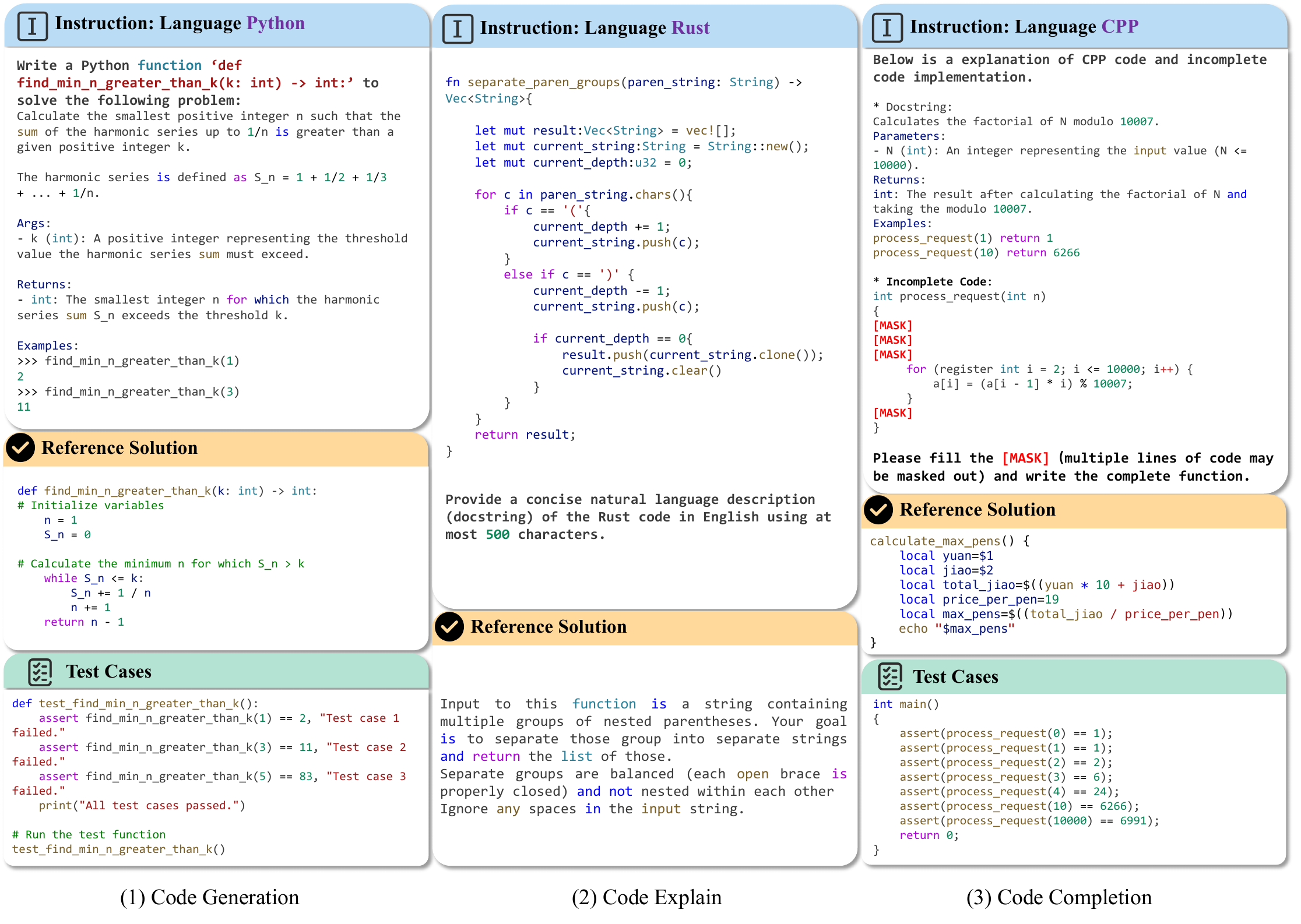}
    \caption{Examples of multilingual code generation, explanation, and completion. }
    \label{fig:bench_cases}
    \vspace{-25pt}
\end{center}
\end{figure*}

\vspace{-10pt}

\subsection{Evaluation Tasks}
\paragraph{Multilingual Code Generation.}
Given the $k$-th programming language $L_{k} \in \{L_{i}\}_{i=1}^{K}$, where $K=40$ is the number of programming languages, we provide the problem description $q^{L_{k}}$ and examples test cases $e^{L_{k}}$ as the input for code LLMs $\mathcal{M}$ to generate the corresponding code $a^{L_{k}}$. We obtain the sampled code result from the code generation distribution $P(a^{L_{k}}|q^{L_{k}}, e^{L_{k}};\mathcal{M})$ from code LLM $\mathcal{M}$, and then feed the test cases into the generated code, where the generated outputs by code should equal the expected outputs. The process can be described as:
\begin{MiddleEquation}
\begin{align}
    r^{L_{k}} = \mathbb{I}(P(a^{L_{k}}|q^{L_{k}}, e^{L_{k}};\mathcal{M}); u^{L_{k}})
    \label{eval_code_generation}
\end{align}
\end{MiddleEquation}where $\mathbb{I}(\cdot)$ is the indicator function by executing the generated code with the given test cases $u^{L_{k}}$. when the generated code $a^{L_{k}}$ passes all test cases, the evaluation result $r=1$, else $r=0$. 

\paragraph{Multilingual Code Explanation.}
To evaluate the understanding capability of code LLMs, we adopt two-pass generation (Code-to-Natural-Language and Natural-Language-to-Code), since the text-similar metrics (e.g. BLEU~\citep{bleu} ) are hindered by the $n$-gram text matching and can not produce an accurate score. We first prompt the code LLMs to generate the natural language description $t^{L_{k}}$ based on the code $a^{L_{k}}$ and then we force the model to restore the original code based on $t^{L_{k}}$. The sampled code from $P(a^{L_{k}}|t^{L_{k}};\mathcal{M})$ is used to evaluate the understanding capability as:
\begin{MiddleEquation}
\begin{align}
    r = \mathbb{I}(P(t^{L_{k}}|a^{L_{k}};\mathcal{M})P(a^{L_{k}}|t^{L_{k}};\mathcal{M}); u^{L_{k}})
    \label{eval_code_explain}
\end{align}
\end{MiddleEquation}where $\mathbb{I}(\cdot)$ is used to check the correctness of the generated code by running the code with test cases.

\paragraph{Multilingual Code Completion.}
Another important scenario is code completion, where the code LLM produces the middle code $a^{L_{k}}_{m}$ based on the prefix code $a^{L_{k}}_{p}$ and suffix code snippet $a^{L_{k}}_{s}$. 
Hence, we concatenate $a^{L_{k}}_{p}$, $a^{L_{k}}_{m}$, and $a^{L_{k}}_{s}$ as the complete code for evaluation as:
\begin{MiddleEquation}
\begin{align}
    r = \mathbb{I}(P(a_{m}^{L_{k}}|a_{p}^{L_{k}},a_{s}^{L_{k}};\mathcal{M}); u^{L_{k}})
    \label{eval_code_completion}
\end{align}
\end{MiddleEquation}where $a^{L_{k}}_{p}$, $a^{L_{k}}_{q}$, and $a^{L_{k}}_{m}$ are concatenated as the complete code to be executed with test cases $u^{L_{k}}$.







\begin{figure*}[ht]
\begin{center}
    \includegraphics[width=0.8\textwidth]{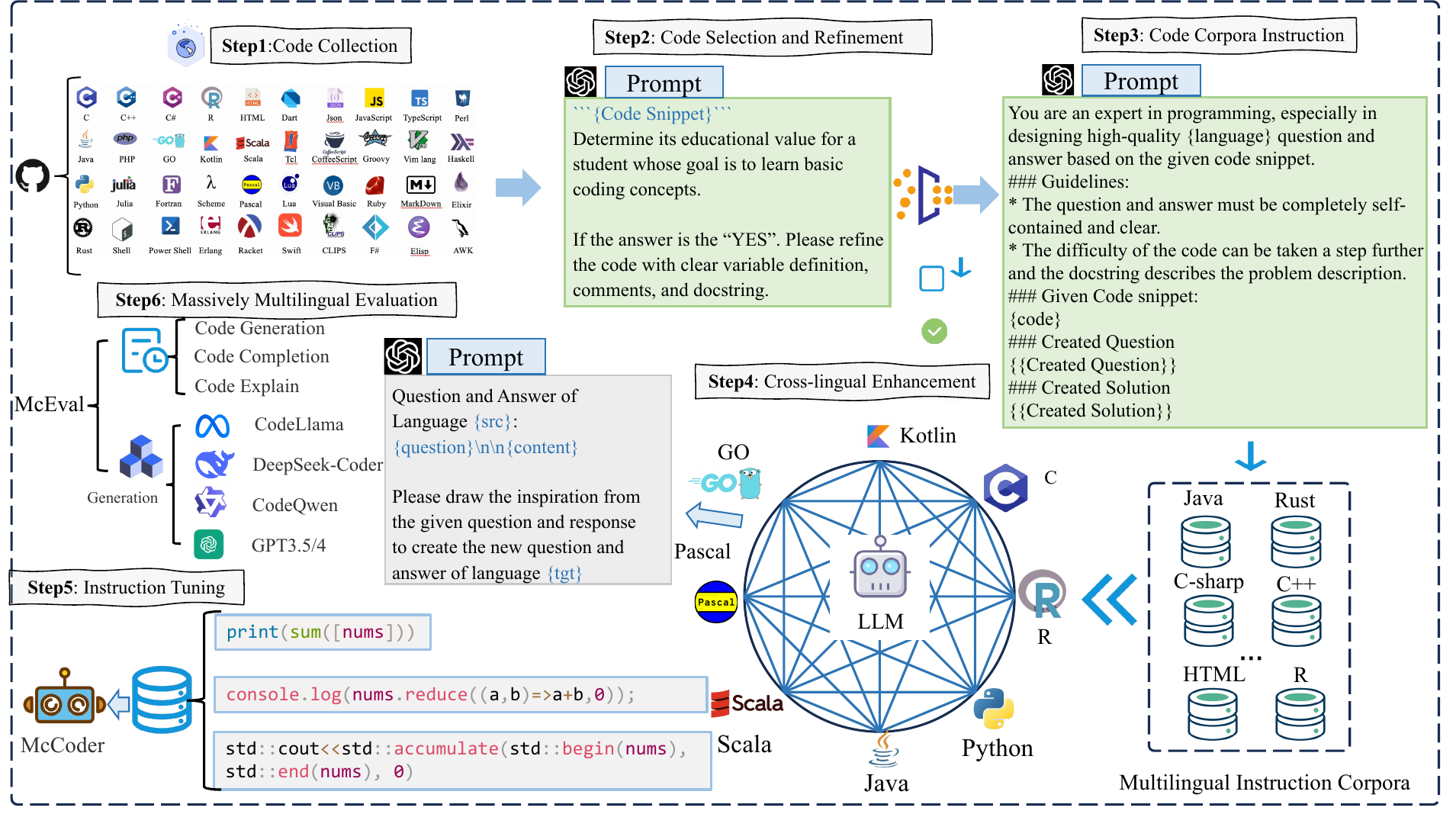}
    \caption{The framework of \ourmethod{}. We first create \instruct{} covering 40 languages from code snippets to fine-tune \ourmethod{}. 20+ existing LLMs and \ourmethod{} are then evaluated on \benchmark{} comprised of multilingual code generation, explanation, and completion.}
    \label{fig:model}
    \vspace{-15pt}
\end{center}
\end{figure*}
\section{\ourmethod{}}
\label{sec:our_method}
\subsection{\instruct{}}
\paragraph{Collection from Code Snippet.}
For a programming language $L_{k}$ ($L_{k} \in \{L_{i}\}_{i=1}^{K}$) and $K$ is the number of programming languages), consider an existing code snippet $c \in D_{c}^{L_{k}}$, we prompt the LLM to select the high-quality code and refine the code to a self-contained code snippet by using the prompt ``\textcolor{blue}{\{Code Snippet\}}\textcolor{darkgray}{\textbackslash nDetermine its educational value for a student whose goal is to learn basic coding concepts.\textbackslash n\textbackslash nIf the answer is `YES'. Please refine the code with clear variable definitions, comments, and docstring.}''. Then, we can obtain the multilingual refined code snippets.
\vspace{-10pt}
\paragraph{Instruction Corpora Generation.}
To construct a comprehensive massively multilingual code instruction corpora $\{D^{L_{i}}\}_{i=1}^{K}$, we prompt the LLMs (\texttt{gpt-4-1106-preview}) to create a problem description  $q^{L_{k}}$ and the corresponding solution $a^{L_{k}}$ by drawing inspiration from the refined code snippet $c^{L_{k}}$. We use LLM to generate instruction dataset by using the prompt ``\textcolor{darkgray}{You are an expert in programming, especially in designing high-quality {language} question and answer based on the given code snippet.\textbackslash n\textbackslash n \#\#\# Guidelines: * The question and answer must be completely self-contained and clear.*\textbackslash n The difficulty of the code can be taken a step further and the docstring describes the problem description.\textbackslash n \#\#\# Given Code snippet: {code}\textbackslash n \#\#\# Created Question {{Created Question}}\textbackslash n \#\#\# Created Solution\textbackslash n {{Created Solution}}}'' in \autoref{fig:model}.
\vspace{-10pt}
\paragraph{Cross-lingual Code Transfer.}
Since the created instruction samples of different programming languages focus on different aspects of coding, we adopt the cross-lingual code transfer to minimize the gap among multiple languages. Given the instruction dataset $D^{L_{i}}$ of language $L_{i}$, we randomly sample a pair $(q^{L_{i}}, a^{L_{i}})$ and force the LLM to modify them to another language $L_{j}$ with a more complex sample $(q^{L_{i} \to L_{j}}, a^{L_{i} \to L_{j}})$. In this way, we can get the derived instruction corpora \{$D^{L_{i} \to L_{j}}\}(i \neq j \land 1 \leq i,j \leq K )$. Finally, we combine $\{D^{L_{k}}\}_{k=1}^{K}$ and $\{D^{L_{i} \to L_{j}}\} (i \neq j \land 1 \leq i,j \leq K )$ as the multilingual instruction corpora \instruct{} $\{D^{L_{k}}_{mc}\}_{k=1}^{K}$ covering 40 programming languages.

\subsection{Multilingual Code Instruction Tuning}
The training objective $\mathcal{L}_{all}$ of the multilingual instruction fine-tuning can be described as:
\begin{MiddleEquation}
\begin{align}
    \mathcal{L}_{all} = -\sum_{k=1}^{K} \mathbb{E}_{q^{L_{k}}, a^{L_{k}} \sim \{ D^{L_{k}} \}_{k=1}^{K} } \left[ \log P(a^{L_{k}}|q^{L_{k}}; \mathcal{M}) \right]
    \label{equ:multilingual_loss}
\end{align}
\end{MiddleEquation}where $q^{L_{k}}$ and $a^{L_{k}}$ are the code-related question and answer from the dataset $D^{L_{k}}$ of language $L_{k}$, respectively. $K$ is the number of programming languages.
\vspace{-10pt}


\section{Experiments}
\subsection{Experiment Setup}
\label{sec:experimental_setup}
\paragraph{Code LLMs.} 

We evaluate 23 models with sizes ranging from 7B to 72B parameters, including general/code LLMs, open/closed-source models, and base/instruction models.
For general models, we evaluate GPTs~\citep{gpt3,gpt4} (GPT-3.5-Turbo, GPT4-Turbo, GPT4-o), Qwen1.5~\citep{Qwen}, Llama3~\citep{llama3}, Phi-3~\citep{phi_3}, and Yi~\citep{yi}.
For code models, we test CodeQwen~\citep{qwen-code-interpreter-eval}, DeepSeekCoder~\citep{deepseek_coder}, CodeLlama~\citep{code_llama}, OCTOCODER~\citep{octopack}, CodeShell~\citep{codeshell}, MagiCoder~\citep{magicoder}, WizardCoder~\citep{wizardcoder}, and Codegemma~\citep{codegemma}.
Furthermore, we fine-tune \ourmethod{} based on CodeQwen1.5 and DeepSeekCoder to further explore the language transfer capabilities of code LLMs.

\vspace{-8pt}
\paragraph{Instruction Corpora.}
The resulting dataset, \instruct{} (110K samples), is comprised of created question-answer pairs and open-source collection~\citep{magicoder}. For supervised fine-tuning (SFT), we utilize CodeQwen-1.5 as the foundational code LLMs. Specifically, we select all Python data from \instruct{}, comprising 50K training samples, for \ourmethod{}-Python training. (Data decontamination by removing exact matches from \benchmark{})


\vspace{-8pt}
\paragraph{Optimization \& Evaluation.}
Our \ourmethod{} based on CodeQwen1.5 are trained for 2 epochs with a cosine scheduler, starting at a learning rate of 2e-5 (3\% warmup steps). We use AdamW~\citep{adamw} as the optimizer and a batch size of 512 (max length $4096$). We adopt the greedy Pass@1 (\%) metric~\citep{kulal2019spoc,codex} for evaluations.
For closed-source LLMs, the answers are generated by the official API. For code explanation, we prompt the LLM to describe the code and then restore the descriptions to the original code. (Details can be found in the Appendix).

\begin{table*}[t]
\centering
\resizebox{1.0\textwidth}{!}{
\begin{tabular}{l|ccccccccccccccccccccc}
\toprule
\textbf{Method}  & Size  & AWK  & C  & C++   & C\#   & Clisp   & Coffee  & Dart   & Elisp   & Elixir   & Erlang  & Fortran   & F\#   & Go   & Groovy   & Haskell & Html & Java  & JS & Json & Julia    \\ \midrule
        GPT-4o (240513) & \faLock{} & \colornum{54.0} & \colornum{60.0} & \colornum{58.0} & \colornum{72.0} & \colornum{60.0} & \colornum{82.0} & \textbf{\colornum{54.9}} & \textbf{\colornum{64.0}} & \bf \colornum{66.0} & \colornum{44.0} & \textbf{\colornum{66.0}} & \textbf{\colornum{78.0}} & \textbf{\colornum{62.0}} & \colornum{80.0} & \textbf{\colornum{90.0}} & \colornum{32.0} & \colornum{81.1} & \textbf{\colornum{62.0}} & \colornum{74.0} & \textbf{\colornum{72.0}} \\ 
        GPT-4~Turbo (231106) & \faLock{} & \textbf{\colornum{70.0}} & \textbf{\colornum{60.0}} & \colornum{64.0} & \textbf{\colornum{80.0}} & \textbf{\colornum{64.0}} & \colornum{72.0} & \colornum{45.1} & \colornum{62.0} & \colornum{56.0} & \colornum{38.0} & \colornum{54.0} & \colornum{74.0} & \colornum{56.0} & \colornum{82.0} & \colornum{78.0} & \colornum{30.0} & \textbf{\colornum{83.0}} & \colornum{60.0} & \colornum{72.0} & \colornum{70.0} \\ 
        GPT-3.5~Turbo (240125) & \faLock{} & \colornum{14.0} & \colornum{54.0} & \colornum{58.0} & \colornum{68.0} & \colornum{54.0} & \colornum{76.0} & \colornum{41.2} & \colornum{26.0} & \colornum{30.0} & \colornum{46.0} & \colornum{40.0} & \colornum{68.0} & \colornum{54.0} & \textbf{\colornum{86.0}} & \colornum{50.0} & \colornum{18.0} & \colornum{71.7} & \colornum{60.0} & \colornum{76.0} & \colornum{54.0} \\
\bottomrule
\end{tabular}}
\caption{Pass@1 (\%) results of different code LLMs for multilingual code generation tasks. ``Avg$_{all}$'' represents the average Pass@1 scores of all code languages on the test set of the \benchmark{}.}
\label{tab:generation}
\vspace{-10pt}
\end{table*}

\begin{table*}[!ht]
\centering
\resizebox{1.0\textwidth}{!}{
\begin{tabular}{l|ccccccccccccccccccccc}
\toprule
\multicolumn{21}{c}{\textit{\textbf{Single-line Completion}}} \\
\midrule
\textbf{Method}  &  Size   & AWK  & C  & C++   & C\#   & Clisp   & Coffee  & Dart   & Elisp   & Elixir   & Erlang  & Fortran   & F\#   & Go   & Groovy   & Haskell & Html & Java  & JS & Json & Julia    \\ \midrule

        GPT-4~Turbo (231106) & \faLock{} & \bf \colornum{92.9} & \bf \colornum{74.4} & \colornum{75.6} & \colornum{89.3} & \bf \colornum{91.1} & \bf \colornum{97.5} & \colornum{76.5} & \bf \colornum{84.2} & \bf \colornum{82.4} & \colornum{54.2} & \colornum{79.6} & \colornum{69.6} & \colornum{81.7} & \colornum{92.6} & \bf \colornum{76.2} & \bf \colornum{57.1} & \colornum{93.9} & \colornum{80.0} & \bf \colornum{93.1} & \colornum{93.3} \\ 
        GPT-3.5~Turbo (240125) & \faLock{} & \colornum{14.3} & \colornum{32.9} & \colornum{20.7} & \colornum{41.7} & \colornum{53.6} & \colornum{68.8} & \colornum{28.6} & \colornum{44.7} & \colornum{19.1} & \colornum{6.3} & \colornum{14.3} & \colornum{37.5} & \colornum{67.1} & \colornum{76.6} & \colornum{11.9} & \colornum{21.4} & \colornum{40.8} & \colornum{35.0} & \colornum{4.2} & \colornum{76.7} \\
        DeepSeekCoder-Instruct & 33B & \colornum{50.0} & \colornum{61.0} & \colornum{72.0} & \colornum{79.8} & \colornum{62.5} & \colornum{78.8} & \colornum{70.4} & \colornum{63.2} & \colornum{66.2} & \colornum{45.8} & \colornum{68.4} & \colornum{60.7} & \colornum{67.1} & \colornum{83.0} & \colornum{38.1} & \colornum{52.4} & \colornum{82.7} & \colornum{65.0} & \colornum{77.8} & \colornum{83.3} \\ 
        OCTOCODER & 16B & \colornum{42.9} & \colornum{47.6} & \colornum{52.4} & \colornum{82.1} & \colornum{35.7} & \colornum{56.3} & \colornum{60.2} & \colornum{34.2} & \colornum{8.8} & \colornum{0.0} & \colornum{0.0} & \colornum{48.2} & \colornum{58.5} & \colornum{73.4} & \colornum{0.0} & \colornum{2.4} & \colornum{81.6} & \colornum{56.3} & \colornum{6.9} & \colornum{0.0} \\
\bottomrule
\end{tabular}}
\caption{Pass@1 (\%) results of different models for multilingual code completion tasks. ``Avg$_{all}$'' represents the average Pass@1 scores of all code languages on the test set of the \benchmark{}.}
\label{tab:completion}
\vspace{-10pt}
\end{table*}

\begin{table*}[t]
\centering
\resizebox{1.0 \textwidth}{!}{
\begin{tabular}{l|ccccccccccccccccccccc}
\toprule
\textbf{Method} &  Size   & AWK  & C  & C++   & C\#   & Clisp   & Coffee  & Dart   & Elisp   & Elixir   & Erlang  & Fortran   & F\#   & Go   & Groovy   & Haskell & Html & Java  & JS & Json & Julia   \\ \midrule
        GPT-4o (240513) & \faLock{} & \colornum{74.0} & \colornum{68.0} & \bf \colornum{72.0} & \colornum{72.0} & \bf \colornum{78.0} & \bf \colornum{76.0} & \bf \colornum{54.9} & \bf \colornum{60.0} & \bf \colornum{66.0} & \colornum{38.0} & \bf \colornum{62.0} & \bf \colornum{78.0} & \bf \colornum{74.0} & \bf \colornum{88.0} & \bf \colornum{86.0} & \bf \colornum{16.0} & \colornum{79.2} & \colornum{68.0} & \colornum{24.0} & \bf \colornum{76.0} \\ 
        GPT-4~Turbo (231106) & \faLock{} & \bf \colornum{84.0} & \bf \colornum{70.0} & \bf \colornum{72.0} & \colornum{52.0} & \colornum{60.0} & \bf \colornum{76.0} & \bf \colornum{54.9} & \colornum{50.0} & \colornum{64.0} & \colornum{42.0} & \colornum{50.0} & \bf \colornum{78.0} & \colornum{68.0} & \bf \colornum{88.0} & \colornum{76.0} & \colornum{10.0} & \colornum{81.1} & \bf \colornum{74.0} & \colornum{18.0} & \colornum{70.0} \\ 
        GPT-3.5~Turbo (240125) & \faLock{} & \colornum{8.0} & \colornum{68.0} & \colornum{64.0} & \colornum{62.0} & \colornum{56.0} & \colornum{66.0} & \colornum{29.4} & \colornum{26.0} & \colornum{42.0} & \colornum{34.0} & \colornum{28.0} & \colornum{56.0} & \colornum{54.0} & \colornum{84.0} & \colornum{54.0} & \colornum{6.0} & \colornum{73.6} & \colornum{58.0} & \colornum{6.0} & \colornum{62.0}\\
        Yi-Large-Turbo & \faLock{} & \colornum{76.0} & \colornum{52.0} & \colornum{48.0} & \colornum{50.0} & \colornum{60.0} & \colornum{60.0} & \colornum{25.5} & \colornum{44.0} & \colornum{42.0} & \colornum{44.0} & \colornum{42.0} & \colornum{60.0} & \colornum{54.0} & \colornum{74.0} & \colornum{68.0} & \colornum{14.0} & \colornum{47.2} & \colornum{48.0} & \colornum{26.0} & \colornum{60.0} \\
\bottomrule
\end{tabular}}
\caption{Pass@1 (\%) results of different models for multilingual code explanation tasks. ``Avg$_{all}$'' represents the average Pass@1 scores of all code languages on the test set of the \benchmark{}.}
\label{tab:explain}
\vspace{-10pt}
\end{table*}

\subsection{Main Results}
\paragraph{Multilingual Code Generation.}
\autoref{tab:generation} shows the Pass@1 results of various models on \benchmark{} for multilingual code generation task. The results reveal a significant disparity between closed-source state-of-the-art models and open-source models across nearly all programming languages. Notably, GPT-4o and GPT-4~Turbo lead the benchmark with substantial performance margins over other models. The \benchmark{} apart from previous benchmarks (such as HumanEval), where various open models have achieved comparable or superior performance. The results indicate that \ourmethod{} exhibits clear improvement over the base model in nearly all the studied programming languages. It is noteworthy that \ourmethod{}, despite being trained with very limited multilingual data, still outperforms other large language models (LLMs) of similar or even larger sizes.

\paragraph{Multilingual Code Explanation.}
\autoref{tab:explain} displays the Pass@1 results for multilingual code explanation tasks. The results show that GPT models still significantly outperform open-source models in the code explanation task. For markup languages (Json and Markdown), the complexity of the code structure makes it difficult to describe accurately in natural language, resulting in generally poorer performance.
Code LLMs need instruction-following capabilities for such complex structures.



\paragraph{Multilingual Code Completion.}
The completion tasks consist of \textit{single-line completion}, \textit{multi-line completion}, \textit{span completion}, and \textit{span completion (light)}. As shown in \autoref{tab:completion}, the Pass@1 results for multilingual code completion tasks indicate that GPT-4~Turbo still achieves the best performance. Additionally, since this task is relatively easier compared to code generation, some open-source models perform comparably to GPT-4~Turbo in certain programming languages.

\section{Further Analysis}
\begin{figure*}[t!]
\begin{center}
    \includegraphics[width=0.7\textwidth]{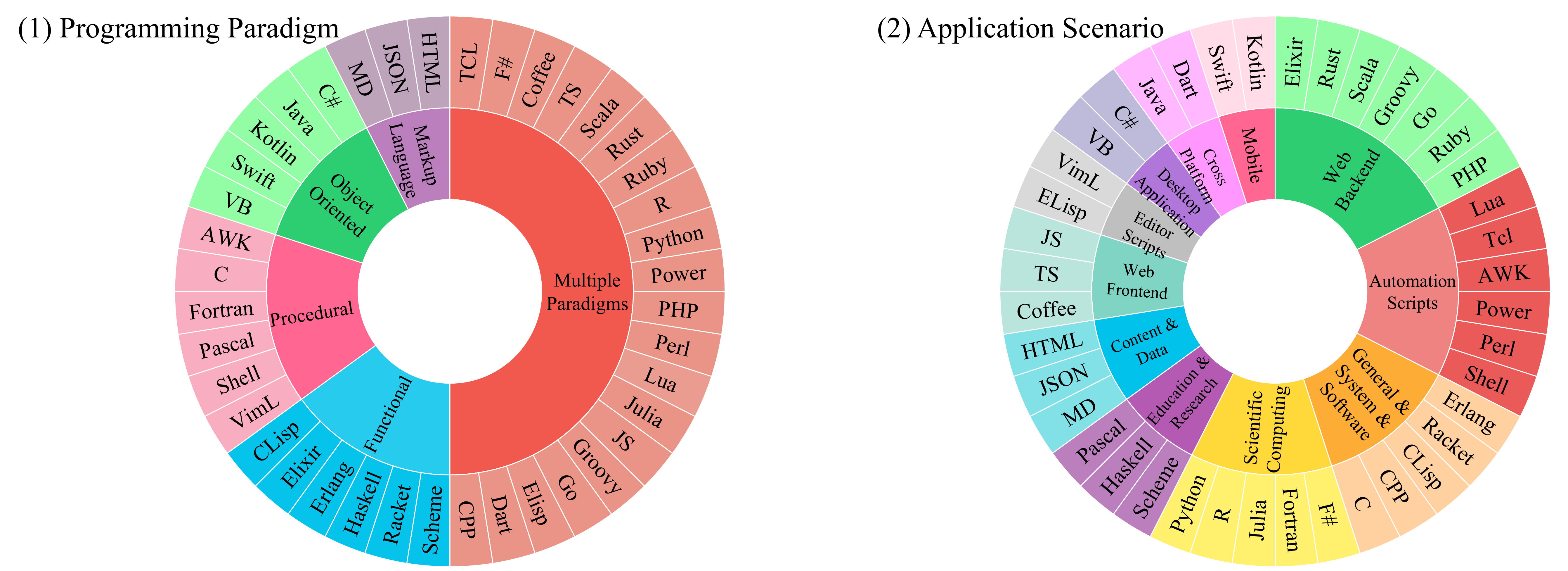}
    \caption{Classification of \benchmark{}. The programming languages in \benchmark{} can be categorized into 5 programming paradigms and 11 application scenarios.}
    \label{fig:code_classes}
    \vspace{-10pt}
\end{center}
\end{figure*}

\begin{figure*}[t!]
\begin{center}
    \includegraphics[width=1.0\textwidth]{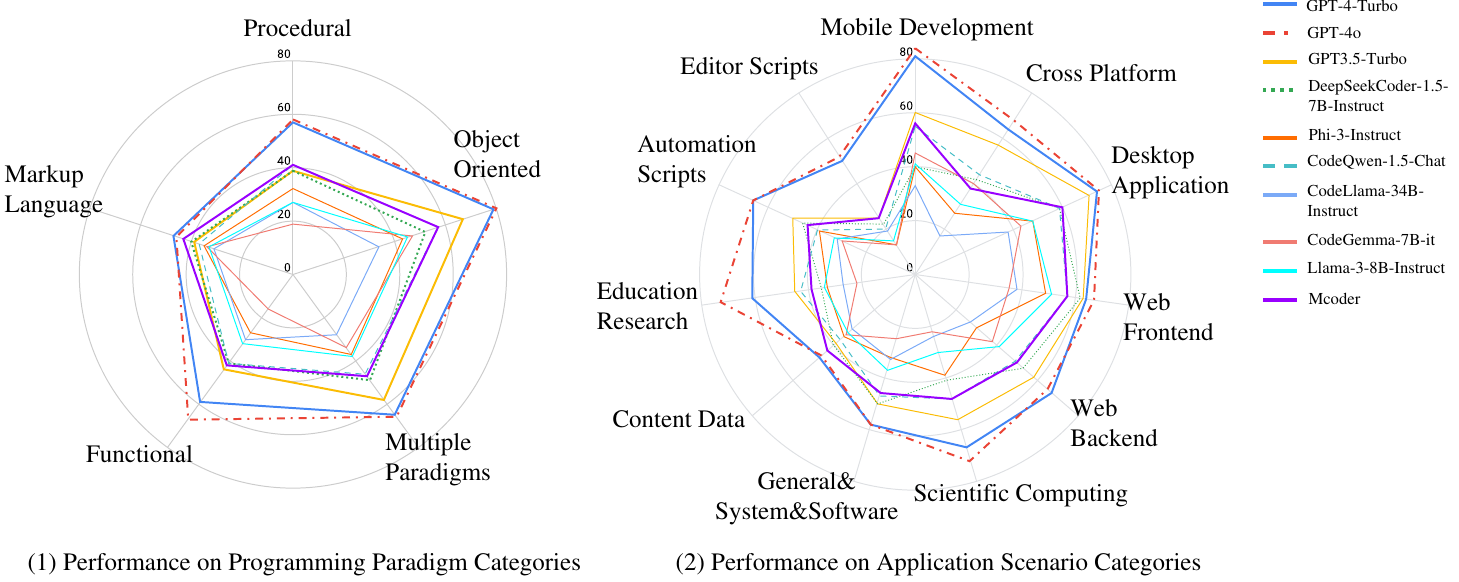}
    \caption{The performance of models in code completion tasks under different categories.}
    \label{fig:code_class_result}
    \vspace{-15pt}
\end{center}
\end{figure*}

\paragraph{Programming Classification.}
In \autoref{fig:code_classes}, we categorize the programming languages of \benchmark{} into 5 programming paradigms and 11 application scenarios and summarize the performance of code LLMs on the code generation task in \autoref{fig:code_class_result}. It can be observed that code LLMs generally perform better in object-oriented and multi-paradigm programming languages (high-resource languages) while performing worse in functional and procedural programming languages (low-resource Languages). In areas like web development and scientific computing, the gap between open-source and closed-source models is narrowing. However, for application scenarios, there is still a substantial gap between open-source models and the closed-source GPT-4 series in low-resource languages related to scripting, mobile development, and educational research. \ourmethod{} performs superior over multiple same-size models and even some larger open-source models.




\begin{figure*}[t!]
\begin{center}
    \includegraphics[width=1.0\textwidth]{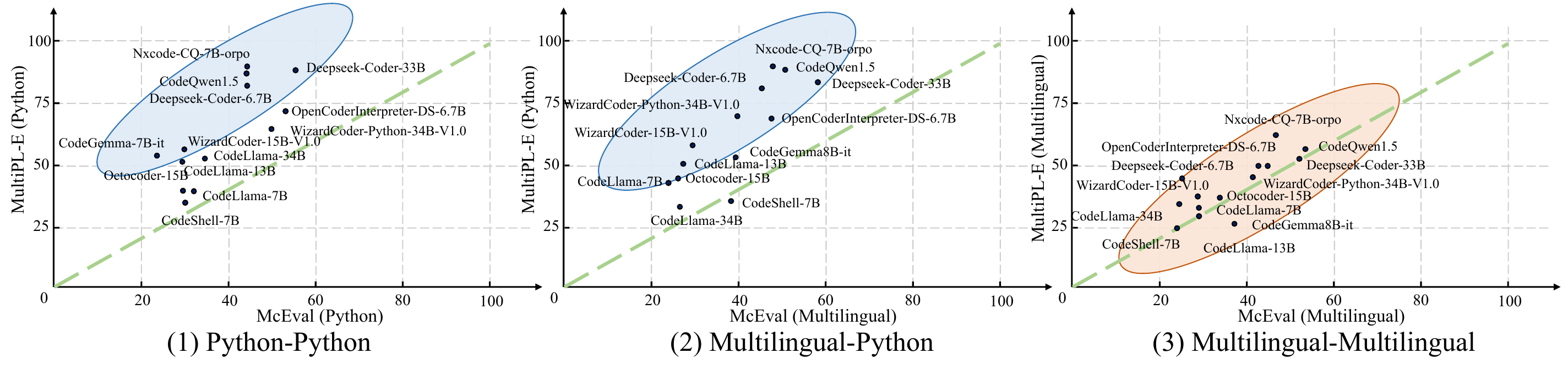}
    \vspace{-5pt}
    \caption{Unbalance performance on different languages across MultiPL-E and our \benchmark{}.}
    \label{fig:cross_dataset}
    \vspace{-25pt}
\end{center}
\end{figure*}

\vspace{-10pt}
\paragraph{Unbalance on Different Languages.}
We compare the results of several open-source models on the MultiPL-E multilingual benchmark with corresponding languages on \benchmark{}. We obtained scores for 11 programming languages (including Python, Java, JavaScript, C++, PHP, Rust, Swift, R, Lua, Racket, Julia) from the BigCode leaderboard.\footnote{\url{https://huggingface.co/spaces/bigcode/bigcode-models-leaderboard}} As shown in \autoref{fig:cross_dataset}(1), due to the simplicity of Python language tasks in this dataset, many models exhibit significant score discrepancies between the two benchmarks. \autoref{fig:cross_dataset}(2) highlights a majority of models within the blue circle, indicating that the current state-of-the-art performance of most models primarily lies in high-resource languages like Python, while their proficiency in low-resource languages awaits further exploration and enhancement. By examining \autoref{fig:cross_dataset}(2) and (3), it becomes evident that all LLMs demonstrate consistent multilingual capabilities between MultiPL-E and \benchmark{}.

\begin{figure*}[t!]
\begin{center}
    \includegraphics[width=0.85\textwidth]{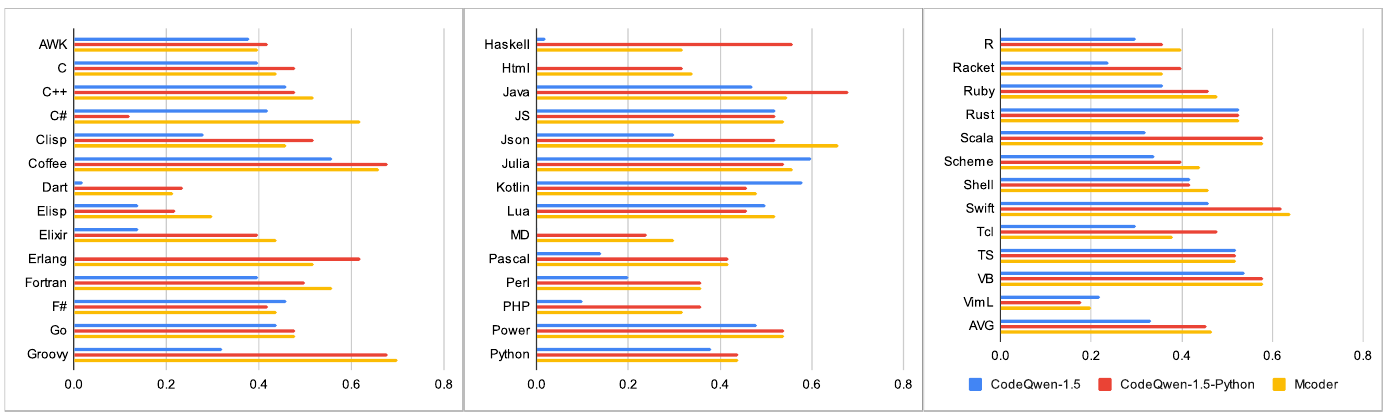}
    \vspace{-5pt}
    \caption{Cross-lingual transferability of LLMs among different languages. We fine-tune CodeQwen-1.5 using Python data in \instruct{} and OSS-Instruct to create CodeQwen-1.5-Python.}
    \label{fig:cross_lingual_transfer}
    \vspace{-10pt}
\end{center}
\end{figure*}

\paragraph{Cross-lingual Transfer.}
We fine-tune the CodeQwen-1.5 model using Python-only data in \instruct{} and compare it with \ourmethod{}.
In \autoref{fig:cross_lingual_transfer}, CodeQwen-1.5 performs well in most high-resource languages, but CodeQwen without alignment exhibits unsatisfactory results in some low-resource languages due to the inability to follow instructions.
As such, with fine-tuning using only Python data, CodeQwen-1.5-Python improves significantly across most languages. It shows that the CodeQwen foundation model already possesses strong coding capabilities but lacks adequate instruction-following skills.
Therefore, fine-tuning with Python-only data can still effectively transfer instruction-following abilities to other languages, resulting in superior multilingual performance.

\paragraph{Difficulty of \benchmark{}.}
Based on algorithmic complexity, we classify \benchmark{} into three levels (Easy/Medium/Hard). In Figure 1, we conduct a statistical analysis of CodeQwen-1.5-Chat's performance on code generation tasks across various languages. For most languages, the code LLM can answer the majority of easy questions but struggles with medium and hard ones.

\begin{figure*}[t!]
\begin{center}
    \includegraphics[width=1.0\textwidth]{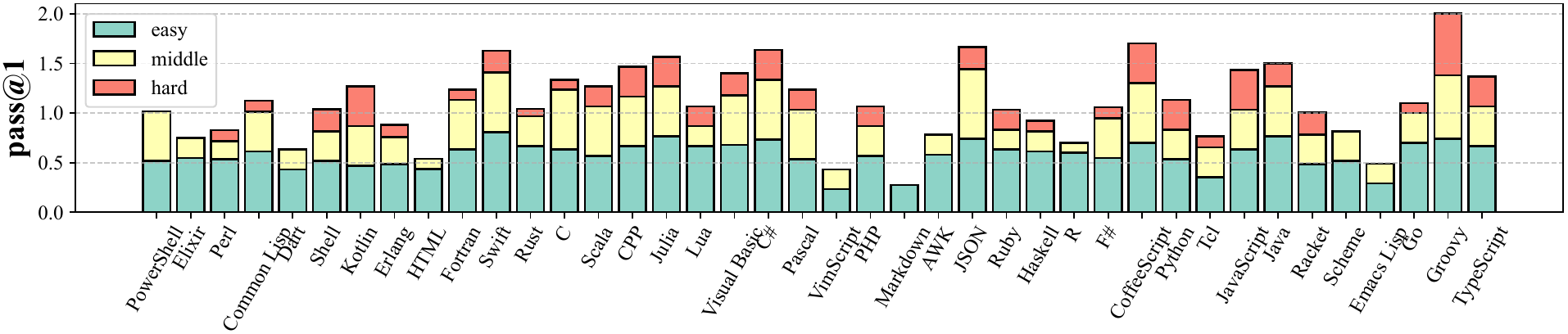}
    \vspace{-5pt}
    \caption{CodeQwen-1.5-Chat performance on \benchmark{} for problems of different difficulty levels.}
    \label{fig:data_difficulty_statistics}
    \vspace{-18pt}
\end{center}
\end{figure*}

\vspace{-3pt}
\section{Related Work}
\vspace{-7pt}
\label{sec:related_work}
For the field of soft engineering, code LLMs~\citep{code_bert,codex,bloom,AlphaCode,santacoder,incoder,codet5,opencodeinterpreter,deepseek_coder} pre-trained on billions of code snippets, such as StarCoder~\citep{starcoder, starcoder2}, CodeLlama~\citep{code_llama}, DeepSeekCoder~\citep{deepseek_coder}, and Code-Qwen~\citep{Qwen}. The development and refinement of code LLMs have been pivotal in automating software development tasks, and supporting code generation/translation/summarization. 

Many benchmarks~\citep{codereval,arcade_nl2code,xcodeeval,BabelCode} have been woven to accurately assess code quality, functionality, and efficiency, such as HumanEval~\citep{codex}, MBPP~\citep{mbpp}, their upgraded version EvalPlus~\citep{evalplus}.
Studies have explored a variety of approaches, ranging from static evaluation using text matching to dynamic methods that involve code execution under a controlled environment. The current benchmarks support code LLMs to evaluate a series of different types of tasks, such as code understanding, code repair~\citep{QuixBugs,debugbench,swe_bench,EvalGPTFix,runbugrun,he2022distribution}, code translation~\citep{CodeTransOcean}. Some works focus on the multilingual scenarios~\citep{odex,mbxp,humaneval_xl,humaneval_x} by extending the Python-only HumanEval/MBPP benchmark (e.g. MultiPL-E~\citep{multipl_e}), which is challenged by the number of the languages and data leaking problem~\citep{livecodebench}.
\vspace{-7pt}

\section{Conclusion}
\label{sec:conclusion}
\vspace{-8pt}
In this work, we push a significant advancement in the assessment of code LLMs by proposing the first massively multilingual code evaluation benchmark (\benchmark{}) by involving an annotation and verification process conducted by professional developers, which spans 40 programming languages and helps comprehensively tackle various tasks, including code generation, explanation, and completion. The multilingual SFT on created instruction corpora \instruct{} further emphasizes the proficiency of LLMs in multiple coding languages.
Systematic evaluations of existing code LLMs on \benchmark{} illuminate the performance disparities among open-source and closed-source models.
Extensive multilingual multitask assessment on \benchmark{} provides a realistic and comprehensive measurement of code LLMs, marking a leap forward for developers utilizing AI techniques to understand and generate code effectively across a wide spectrum of programming languages.


\bibliography{reference}
\bibliographystyle{natbib}

\clearpage
\appendix






\section{Data Annotation}
\subsection{Human Annotation}
To create the massively multilingual code evaluation benchmark \benchmark{}, the annotation of multilingual code samples is conducted utilizing a comprehensive and systematic human annotation procedure, underpinned by rigorously defined guidelines to ensure accuracy and consistency. Initially, 10 software developers in computer science are recruited as multilingual programming annotators with proven proficiency in the respective programming languages. 

Following a detailed training session on the annotation protocol, annotators are tasked with creating problem definitions and the corresponding solution.

The guidelines for our annotation training session primarily cover the following aspects:
\begin{itemize}
\item \textbf{Standardized Format:} We provide an annotation example for 40 programming languages. Annotators are required to adhere to this standardized format when annotating data.
\item \textbf{Accessibility:} The reference data for our annotations is sourced from materials available under permissive licenses, allowing unrestricted use and redistribution for research purposes.
\item \textbf{Difficulty Level:} We provide annotators with detailed guidelines on the difficulty classification for each language. Annotators must strictly follow these guidelines to label problems according to their respective difficulty levels based on algorithmic complexity and functionality.
\item \textbf{Self-Contained:} Annotators are required to thoroughly review their annotated problems to ensure that the problem descriptions include all necessary information for solving them without ambiguity. The provided example inputs and outputs must be correct, the reference answers must execute correctly, and the test cases written should comprehensively evaluate the accuracy of the functions.
\end{itemize}

In addition, we encourage annotators to use GPT-4~Turbo to assist in constructing draft questions. \autoref{tab:anno_prompt} shows a one-shot draft question generation prompt. In the Reference question section, the user enters the question description that they have selected or written, and then GPT-4 quickly generates a unified format of question description, question answers, and test examples. Annotators need to conduct a comprehensive review and modification of the draft questions generated by GPT-4 to ensure the correctness of the annotations.

\begin{table}[!ht]
    \centering
    \scalebox{0.95}{
\begin{promptbox}[Question Generation Prompt]{lightblue}
\textbf{\#\#\# You are a coding expert who specializes in designing programming-related exam questions.}\\
Your task now is to be given a language type and related examples. You generate the corresponding function, docstring, corresponding solution, and corresponding test cases based on the provided references.\\
\textbf{\#\#\# The following is an example using the programming language Kotlin:}\\
\textbf{\#\#\# Function declaration(give the func name, args and return arg if necessary.)}\\
fun hasCloseElements(numbers: List<Double>, threshold: Double): Boolean \\
\textbf{\#\#\# Docstring(describe the question in English and give some cases)}  \\
/** \\
    * Check if, in given list of numbers, any two numbers are closer to each other than \\
    * given threshold. \\
    Example: \\
    >>> hasCloseElements(listOf(1.0, 2.0, 3.9, 4.0, 5.0, 2.2), 0.3) \\
        true \\
    >>> hasCloseElements(listOf(1.0, 2.0, 3.9, 4.0, 5.0, 2.2), 0.05) \\
        false \\
**/ \\
\textbf{\#\#\# Think about this Question:} \\
To check if there are any two numbers in a given list that are closer to each other than a given threshold in Kotlin, you can follow these steps: \\
Sort the list of numbers in ascending order. \\
Iterate through the sorted list and compare the difference between adjacent numbers. \\
If the difference between any two adjacent numbers is less than the given threshold, then return true, indicating that there are two numbers closer to each other than the threshold. \\
If no such pair is found after iterating through the entire list, return false. \\

\textbf{\#\#\# Solution in programming language Kotlin:} \\
\{ \\
for ((idx, elem) in numbers.withIndex()) \{ \\
~~~~~~~~for ((idx2, elem2) in numbers.withIndex()) \{ \\
~~~~~~~~~~~~if (idx != idx2) \{ \\
~~~~~~~~~~~~~~~~val distance = Math.abs(elem - elem2) \\
~~~~~~~~~~~~~~~~if (distance < threshold) \{ \\
~~~~~~~~~~~~~~~~~~~~return true \\
~~~~~~~~~~~~~~~~\}\}\} \\
~~~~return false \} \\

\textbf{\#\#\# Corresponding test cases:} \\
    check(hasCloseElements(listOf(1.0, 2.0, 3.9, 4.0, 5.0, 2.2), 0.3) == true) \\
    check(hasCloseElements(listOf(1.0, 2.0, 3.9, 4.0, 5.0, 2.2), 0.05) == false) \\
    ...
    \\
\textbf{\#\#\# Reference question:} \\
A laboratory has N test tubes containing chemicals, arranged in a straight line. If chemicals are placed in M consecutive test tubes, an explosion will occur, so some test tubes may not contain chemicals.
Task: For a given N and M, find the total number of ways to place chemicals without causing an explosion\\

\textbf{\#\#\# Base on the reference question above, please use language \"Kotlin\", follow below steps:} \\
1. Generate "Func declaration", give the func name, args, and return arg if necessary. \\
2. Generate "Docstring", describe the question in English, and give cases to explain.\\
3. Think about this question.\\
4. Generate "Solution".\\
5. Generate corresponding "Test cases". \\

Please give an answer that is as accurate as possible. \\

\end{promptbox}
    }
    \vspace{5pt}
    \caption{A one-shot draft question generation prompt.}
    \label{tab:anno_prompt}
\end{table}






We paid all the annotators the equivalent of \$6 per question and provided them with a comfortable working environment, free meals, and souvenirs. We also provided the computer equipment and GPT-4 interface required for labeling. We labeled about 2,000 questions in total and employed them to check the quality of the questions/answers, and the total cost was about \$12,000 in US dollars. The annotators checked the derived tasks, including multilingual code explanation and code completion.

\subsection{Quality Control}
A dual-pass system is adopted to ensure the correctness of the created benchmark \benchmark{}, where each snippet is independently annotated by at least two annotators to minimize subjective bias and errors. Discrepancies between annotators are resolved through consensus or adjudication by a senior annotator. This meticulous annotation process ensures the creation of a high-quality, multilingual programming dataset that facilitates the comprehensive analysis and understanding of code examples across different languages. Finally, three volunteers are employed to evaluate whether the correctness of the benchmark \benchmark{} can reach 90\% accuracy and then correct the errors.

\section{Statistic Details}

In \autoref{tab:detail_data}, we display the number of questions, test cases, and difficulty level corresponding to the three tasks in \benchmark{}, as well as the number of questions in the four sub-tasks of the completion task.
Among these tasks, the \textit{span completion (light)} task is similar in form to the \textit{span completion} task. However, in the \textit{span completion (light)} task, each problem is paired with all the corresponding code, making it a balanced version of the \textit{span completion} task (fewer samples for fast inference and the same test size of each programming language). The results of \textit{span completion (light)} can better reflect the differences in model performance across different languages.
\begin{table*}[ht]
    \centering
    \resizebox{0.85 \textwidth}{!}{
    \begin{tabular}{l|l|c|c|c}
    \toprule
Task& Sub Task  & \#Questions & \#Test Cases & \#Difficulty Level (Easy/Medium/Hard)\\
\midrule 
Code Generation & - & 2007 &  \multirow{6}{*}{10086}&  \multirow{6}{*}{1221~/~401~/~385}\\
Code Explanation & - & 2007 & \\
\multirow{4}{*}{Code Completion} & Single-Line  & 2998  &\\
 & Multi-Line & 2998 &\\
 & Span & 4014 &\\
 & Span (Light) & 2007 &\\

\bottomrule
\end{tabular}
}
\caption{Statistic details of \benchmark{}}
\label{tab:detail_data}
\end{table*}

In \autoref{tab:compare_bench}, 
We compared \benchmark{} with other multilingual benchmarks. It is noteworthy that our benchmark provides a significant supplement to current benchmarks in terms of both the variety of programming languages and the number of questions.

\begin{table*}[ht]
    \centering
    \resizebox{0.85 \textwidth}{!}{
    \begin{tabular}{l|cccccc}
    \toprule
Benchmark &  Multi Task &\#Program Lang. & Data source & \#Questions\\
\midrule 
MuliPL-E    & \ding{55} & 18  & Translate &  \textasciitilde 3,000\\ 
MBXP        &  \ding{51} & 10  & Translate &  12,425 \\
HumanEval-X &  \ding{51}  & 5    & Hand-Written & 820 \\
HumanEval-XL &\ding{55}  & 12  & Hand-Written & 22,080\\
\benchmark{}&  \ding{51}  & 40  & Hand-Written &  16,031\\
\bottomrule
\end{tabular}
}
\caption{Comparison between \benchmark{} and other multilingual code benchmarks.}
\label{tab:compare_bench}
\end{table*}

\section{Examples in \benchmark{}}

\begin{figure*}[!ht]
\begin{center}
    \includegraphics[width=1.0\textwidth]{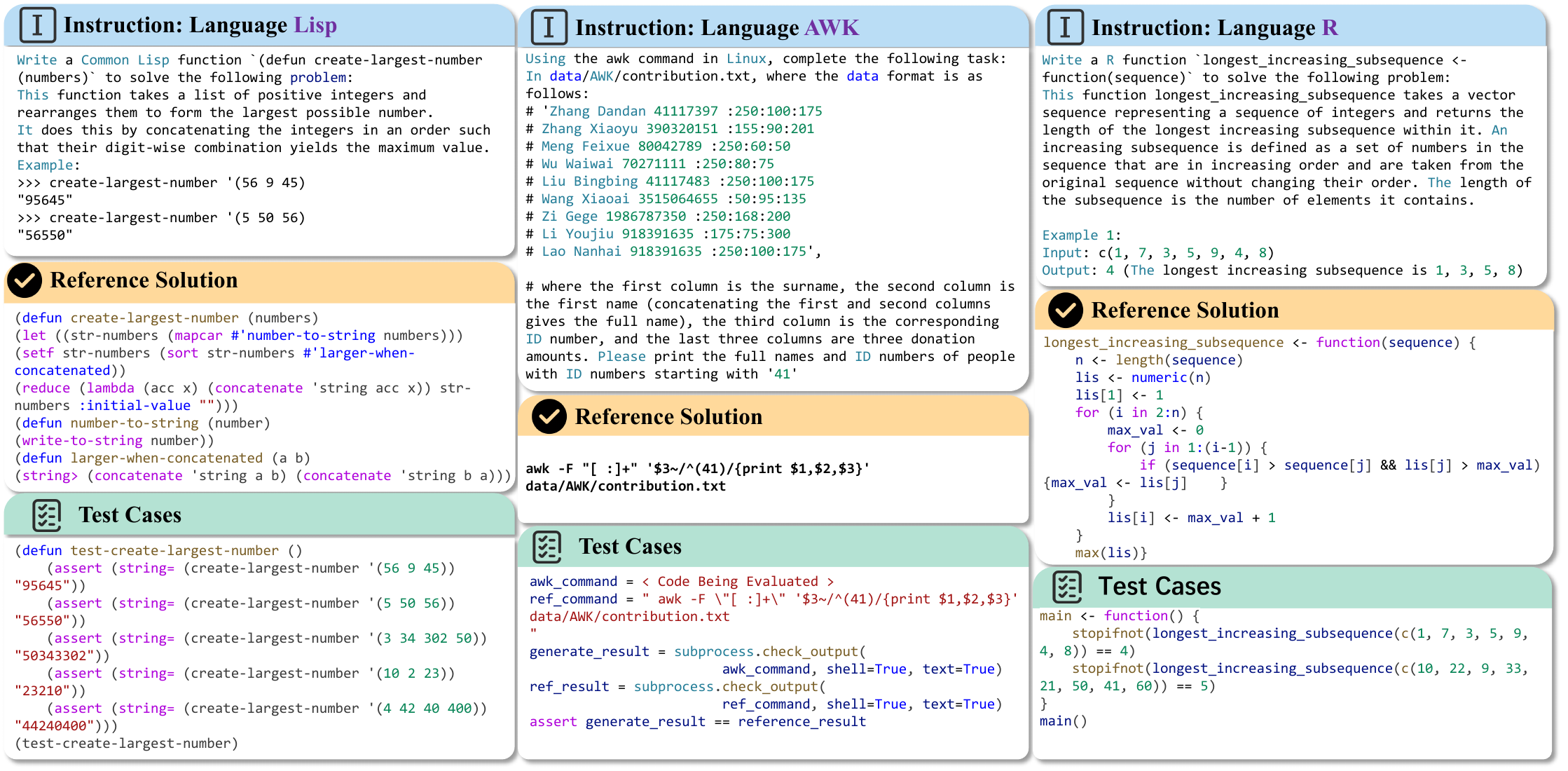}
    \caption{Examples of multilingual generation. The data mainly consists of an instruction part (including function name, function description, and function call cases), a reference solution, and a test cases part. \textbf{Left.}~Shows an example of the Lisp language. \textbf{Middle.}~Shows a file processing programming task in AWK language. During the evaluation, the corresponding file processing result by the generated code will be compared with the reference answer. \textbf{Right.}~Shows an example of the R language.}
    \label{fig:generation_example}
    \vspace{-10pt}
\end{center}
\end{figure*}

\autoref{fig:generation_example} display three examples of multilingual generation. 

\begin{figure*}[!ht]
\begin{center}
    \includegraphics[width=1.0\textwidth]{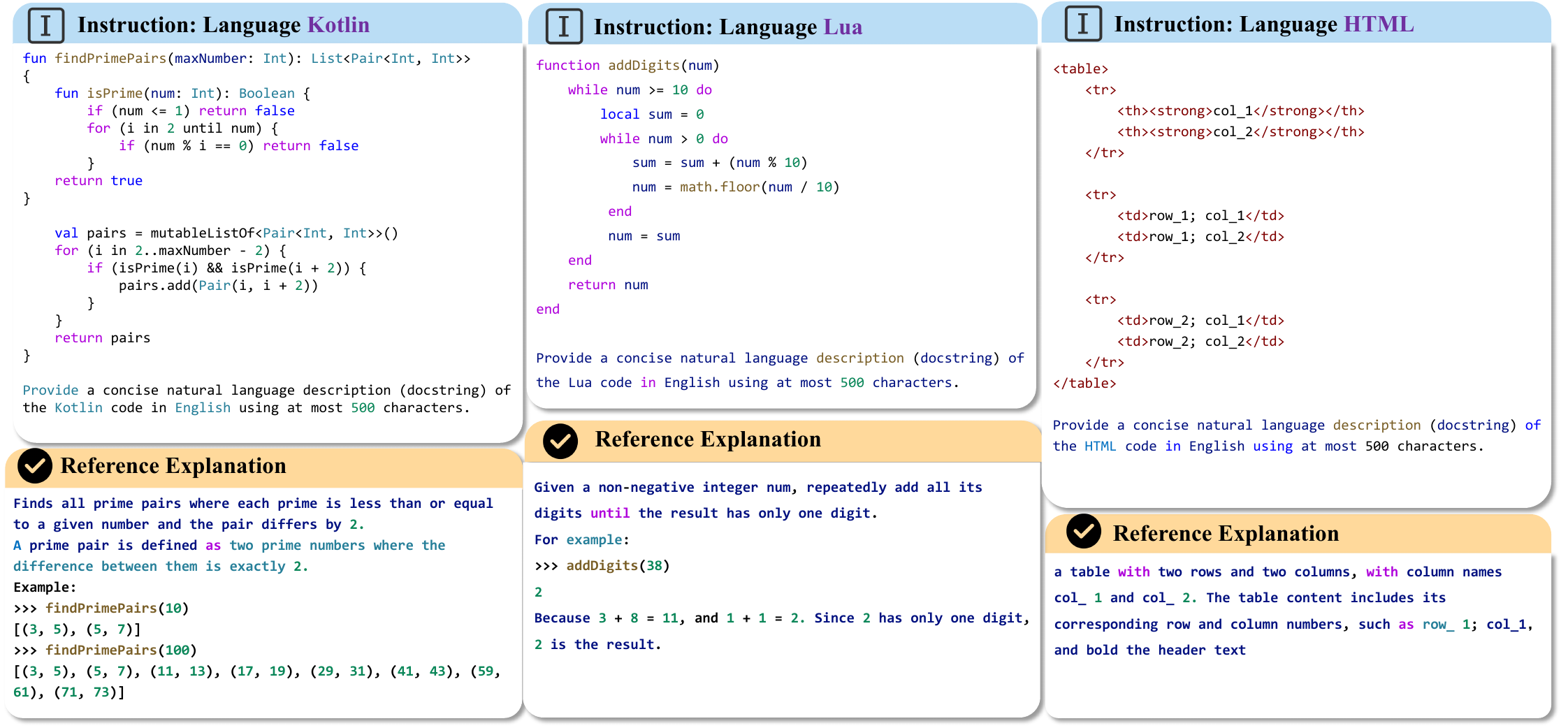}
    \caption{Examples of multilingual explanation. The data mainly consists of an instruction part (including a complete function), a reference Explanation. \textbf{Left.}~Shows an example of the Kotlin language. \textbf{Middle.}~Shows an example of the Lua language. \textbf{Right.}~Shows an example of the HTML language.}
    \label{fig:explain_example}
    \vspace{-10pt}
\end{center}
\end{figure*}

In \autoref{fig:explain_example}, we show three examples of multilingual explanation. 

\begin{figure*}[!ht]
\begin{center}
    \includegraphics[width=1.0\textwidth]{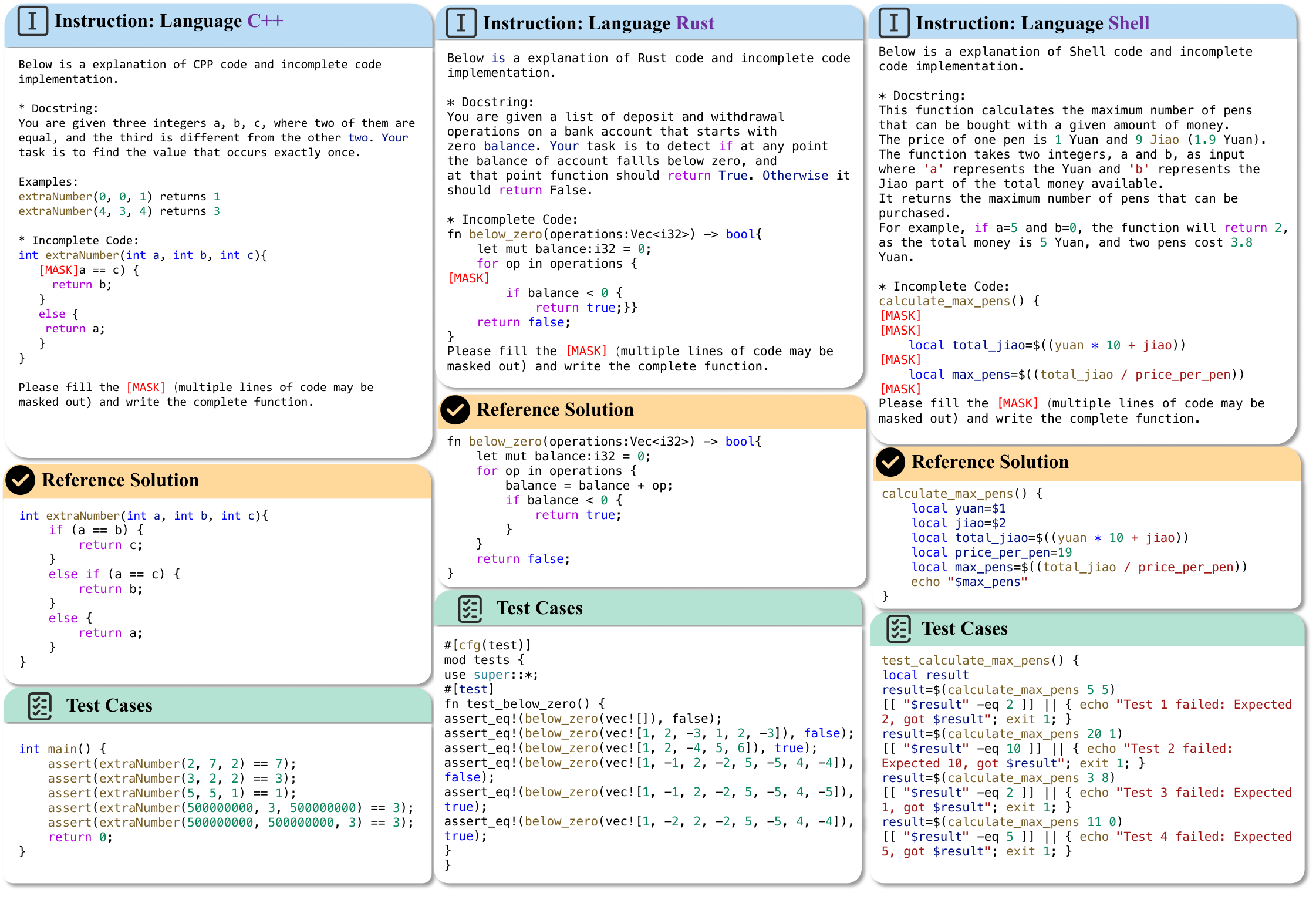}
    \caption{Examples of multilingual completion. The data mainly consists of an instruction part (including an incomplete function ), a reference complete code solution, and test cases. \textbf{Left.}~Shows an span completion  example of the C++ language. \textbf{Middle.}~Shows a single-line completion example of the Rust language. \textbf{Right.}~Shows a multiple-line completion example of the Shell language.}
    \label{fig:completion_example}
    \vspace{-10pt}
\end{center}
\end{figure*}

In \autoref{fig:completion_example}, we display three examples of multilingual explanation. The three examples from left to right correspond to the span completion task, the single-line completion task, and the multi-line completion task.

\section{Evaluation}

\begin{figure*}[!ht]
\begin{center}
    \includegraphics[width=1.0\textwidth]{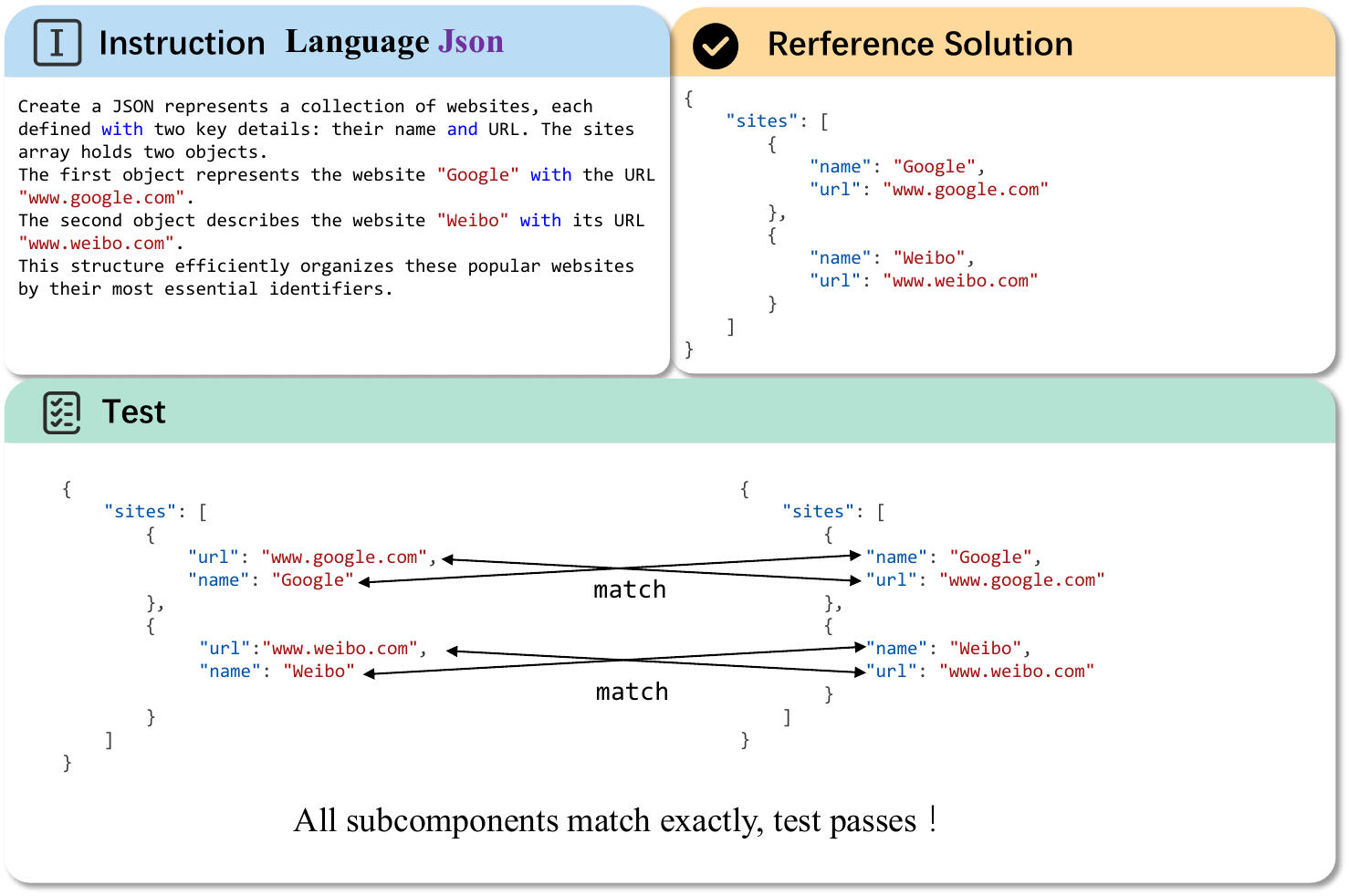}
    \caption{Examples of Markup language (Json) generation task evaluation. }
    \label{fig:json_eval_example}
    \vspace{-10pt}
\end{center}
\end{figure*}

For programming languages other than markup languages, we use an execution-based correctness metric by running the code with the provided test cases. For markup languages, we use the Exact Match metric for evaluation. Taking Json as an example, we parse all subcomponents in Json. If the model result is exactly the same as the subcomponent of the reference solution, the model generation result is considered correct. An example of Markup language (Json) is shown in \autoref{fig:json_eval_example}.

We adopt the greedy Pass@1 (\%) metric~\citep{kulal2019spoc,codex} for our evaluations.
For closed-source models, we generate answers through the official API service. For open-source models, we prioritize using vLLM~\citep{pageattention} for faster inference if the model is supported by vLLM. Otherwise, we perform inference with the Distributed Data Parallel (DDP) module from PyTorch.
For the code generation and code completion tasks, we extract the functional part of the code from the model outputs and combine it with corresponding test cases to form compilable and executable code. For the code explanation task, we adopt a two-pass generation approach (Code-to-Natural-Language and Natural-Language-to-Code). The extraction and execution process for this task is consistent with the previous two tasks.
We conduct all evaluations in a Docker environment. Detailed information on the code compilation and execution environment are displayed in \autoref{tab:runtime_env}.  We have uploaded the Docker image to the Docker hub to facilitate the reproduction of results and the evaluation of new models.

\begin{table}[htbp]
    \centering
\resizebox{0.95 \textwidth}{!}{
    \begin{tabular}{l|l}
    \toprule
    Language & Runtime Environments\\
    \midrule 
    AWK & GNU bash, version 4.4.20(1)-release (x86\_64-pc-linux-gnu) \\ 
        C & gcc (Ubuntu 7.5.0-3ubuntu1\~18.04) 7.5.0  \\ 
        C\# & dotnet 8.0.100 \\ 
        CPP & g++ (Ubuntu 7.5.0-3ubuntu1\~18.04) 7.5.0 \\ 
        CoffeeScript & CoffeeScript version 1.12.7 \\ 
        Common Lisp & SBCL 1.4.5.debian \\ 
        Dart & Dart SDK version: 3.3.1 (stable) \\ 
        Elixir & elixir 1.3.3 \\ 
        Emacs Lisp & GNU Emacs 25.2.2 \\ 
        Erlang & Erlang/OTP 20 [erts-9.2] \\ 
        F\# & dotnet 8.0.100 \\ 
        Fortran & GNU Fortran (Ubuntu 7.5.0-3ubuntu1\~18.04) 7.5.0 \\ 
        Go & go version go1.18.4 linux/amd64 \\ 
        Groovy & Groovy Version: 4.0.16 JVM: 17.0.9 Vendor: Oracle Corporation OS: Linux \\ 
        HTML & - \\ 
        Haskell & The Glorious Glasgow Haskell Compilation System, version 9.4.7 \\ 
        Json & - \\ 
        Java & javac 11.0.19 \\ 
        JavaScript & Node.js v16.14.0 \\ 
        Julia & julia v1.9.4 \\ 
        Kotlin & kotlinc-jvm 1.9.21 (JRE 17.0.9+11-LTS-201) \\ 
        Lua & Lua 5.4.6  Copyright (C) 1994-2023 Lua.org, PUC-Rio \\ 
        Markdown & - \\ 
        PHP & PHP 7.2.24-0ubuntu0.18.04.17 (cli) (built: Feb 23 2023 13:29:25) ( NTS ) \\ 
        Pascal & Free Pascal Compiler version 3.2.2 [2021/05/16] for x86\_64 \\ 
        Perl & perl 5, version 26, subversion 1 (v5.26.1) built for x86\_64-linux-gnu-thread-multi \\ 
        PowerShell & PowerShell 7.4.0 \\ 
        Python & Python 3.8.12 \\ 
        R & R version 3.4.4 \\ 
        Racket & Racket v6.11 \\ 
        Ruby & ruby 2.5.1p57 (2018-03-29 revision 63029) [x86\_64-linux-gnu] \\ 
        Rust & rustc 1.74.0 (79e9716c9 2023-11-13) \\ 
        Scala & Scala code runner version 3.3.1 -- Copyright 2002-2023, LAMP/EPFL \\ 
        Scheme & Racket v6.11 \\ 
        Shell & GNU bash, version 4.4.20(1)-release (x86\_64-pc-linux-gnu) \\ 
        Swift & Swift version 5.9.2 (swift-5.9.2-RELEASE) \\ 
        Tcl & tclsh 8.6.11 \\ 
        TypeScript & tsc Version 5.3.3 \\ 
        VimScript & VIM - Vi IMproved 9.0 (2022 Jun 28, compiled Dec 20 2023 18:57:50) \\ 
        Visual Basic & dotnet 8.0.100 \\ 
    \bottomrule
    \end{tabular}
    }
    \vspace{5pt}
    \caption{Runtime environments for different programming languages.}
    \label{tab:runtime_env}
\end{table}

\subsection{Optimization Details}
All \ourmethod{} models are fine-tuned using 8 NVIDIA A800-80GB GPUs. The models are trained for 2 epochs with a cosine scheduler, starting at a learning rate of 2e-5 and incorporating a 3\% warmup phase. Training a model takes about 5 hours. We used AdamW~\citep{adamw} as the optimizer and a batch size of 512 with a sequence truncation length of 4096. We use PyTorch's Fully Sharded Data Parallel (FSDP) to perform distributed training of the model and use gradient checkpointing technology and gradient accumulation to save memory and achieve training with a larger batch size.

\section{Extra Results}
\subsection{Programming Classification}
As shown in \autoref{tab:program_diagram_result} and \autoref{tab:app_generation_result}, we comprehensively display the code generation performance of the models we tested across various programming paradigms and application scenarios.
 \begin{table*}[htbp]
    \centering
    \resizebox{1.00 \textwidth}{!}{
    \begin{tabular}{l|ccccc}
    \toprule
        Method & Procedural & Object Oriented & Multiple Paradigms & Functional & Markup Language \\ 
        \midrule
        GPT-4o (240517) & \textbf{58.0} & \textbf{79.8} & \textbf{65.9} & \textbf{67.0} & 46.0 \\ 
        GPT-4~Turbo (231106) & 56.7 & 78.7 & 65.2 & 59.3 & \textbf{46.7} \\ 
        GPT-3.5-Turbo (240125) & 38.7 & 66.8 & 57.6 & 44.3 & 39.3 \\ 
        Codegemma-7b-it & 19.3 & 46.6 & 34.0 & 16.3 & 34.0 \\ 
        CodeLlama-13b-Instruct & 21.3 & 32.0 & 27.0 & 32.3 & 28.0 \\ 
        CodeLlama-34b-Instruct & 27.3 & 33.6 & 28.0 & 30.0 & 30.7 \\ 
        CodeLlama-7b & 20.3 & 28.1 & 23.4 & 26.7 & 30.7 \\ 
        CodeQwen-1.5-7b-Chat & 41.3 & 57.3 & 46.3 & 41.0 & 37.3 \\ 
        Codeshell-7b-chat & 16.0 & 24.1 & 25.7 & 14.0 & 34.7 \\ 
        Codestral-22B-v0.1 & 40.0 & 67.6 & 54.1 & 39.7 & 40.7 \\ 
        DeepSeekCoder-33b-instruct & 52.7 & 62.8 & 56.3 & 52.0 & 34.7 \\ 
        DeepSeekCoder-1.5-7b-instruct & 39.0 & 51.8 & 48.8 & 41.0 & 40.0 \\ 
        Magicoder-S-DS-6.7B & 45.7 & 58.5 & 49.4 & 49.0 & 32.0 \\ 
        Llama-3-8B-Instruct & 27.3 & 44.7 & 38.0 & 32.0 & 33.3 \\ 
        Nxcode-CQ-7B-orpo & 40.7 & 54.9 & 45.5 & 41.3 & 36.7 \\ 
        OCTOCODER & 20.7 & 28.9 & 21.9 & 25.0 & 25.3 \\ 
        OpenCodeInterpreter-DS-6.7B & 40.7 & 57.7 & 46.4 & 42.0 & 42.0 \\ 
        Phi-3-medium-4k-instruct & 32.3 & 43.1 & 36.6 & 26.7 & 35.3 \\ 
        Qwen1.5-72B-Chat & 38.3 & 37.2 & 36.2 & 29.3 & 39.3 \\ 
        WizardCoder-15B-V1.0 & 19.0 & 31.6 & 34.2 & 24.0 & 6.7 \\ 
        WizardCoder-Python-34B & 27.7 & 43.9 & 38.2 & 33.7 & 36.0 \\ 
        \ourmethod{} & 41.3 & 57.3 & 47.4 & 42.3 & 43.3 \\ 
\bottomrule
\end{tabular}}

\caption{Pass@1(\%) results of code generation performance across various programming paradigms}
\label{tab:program_diagram_result}
\end{table*}

\begin{table*}[htbp]
    \centering
    \resizebox{1.00 \textwidth}{!}{
    \begin{tabular}{l|lllllllllll}
    \toprule
        Method & Mobile & Cross & Desktop & Frontend & Backend & Scientific & General & Content & Education & Scripts & Editor \\ 
    \midrule
        GPT-4o (230517) & \textbf{84.0} & \textbf{68.3} & \textbf{75.0} & \textbf{66.7} & 64.6 & \textbf{71.6} & \textbf{57.6} & 46.0 & \textbf{72.7} & \textbf{65.7} & \textbf{52.0} \\ 
        GPT-4 Turbo (231106) & 81.0 & 64.4 & 74.0 & 64.0 & \textbf{66.6} & 66.8 & \textbf{57.6}& \textbf{46.7} & 60.7 & \textbf{65.7} & 50.0 \\ 
        GPT-3.5 (240125) & 60.0 & 56.7 & 71.0 & 63.3 & 57.5 & 55.6 & 50.4 & 39.3 & 45.3 & 50.0 & 25.0 \\ 
        Codegemma-7b-it & 45.0 & 40.4 & 43.0 & 34.7 & 37.7 & 21.6 & 24.8 & 34.0 & 22.0 & 29.7 & 13.0 \\ 
        code-Llama-13b & 30.0 & 15.4 & 39.0 & 34.7 & 28.0 & 23.2 & 34.8 & 28.0 & 24.0 & 27.7 & 13.0 \\ 
        CodeLlama-34b-Instruct & 33.0 & 17.3 & 38.0 & 38.0 & 27.2 & 24.0 & 32.8 & 30.7 & 26.7 & 31.7 & 19.0 \\ 
        Code-Llama-7b-Instruct & 24.0 & 12.5 & 37.0 & 29.3 & 22.7 & 20.8 & 29.2 & 30.7 & 19.3 & 27.0 & 14.0 \\ 
        CodeQwen-1.5-7b & 55.0 & 44.2 & 59.0 & 56.7 & 48.7 & 47.6 & 46.8 & 37.3 & 42.7 & 40.0 & 20.0 \\ 
        Codeshell-7b-chat & 23.0 & 14.4 & 26.0 & 40.7 & 26.1 & 17.2 & 21.2 & 34.7 & 13.3 & 22.7 & 8.0 \\ 
        Codestral-22B-v0.1 & 68.0 & 58.7 & 64.0 & 57.3 & 55.0 & 54.0 & 44.8 & 40.7 & 30.0 & 53.3 & 28.0 \\ 
        DeepSeekCoder-33b-instruct & 63.0 & 50.0 & 57.0 & 68.0 & 60.6 & 54.8 & 54.0 & 34.7 & 56.7 & 52.7 & 35.0 \\ 
        DeepSeekCoder-1.5-7b-instruct & 40.0 & 42.3 & 59.0 & 62.0 & 52.7 & 40.8 & 50.0 & 40.0 & 34.7 & 46.0 & 22.0 \\ 
        Magicoder-S-DS-6.7B & 49.0 & 43.3 & 64.0 & 60.7 & 50.4 & 49.6 & 52.4 & 32.0 & 48.7 & 49.7 & 24.0 \\ 
        Llama-3-8B-Instruct & 41.0 & 30.8 & 48.0 & 50.7 & 40.5 & 30.0 & 37.2 & 33.3 & 34.0 & 33.0 & 15.0 \\ 
        Nxcode-CQ-7B-orpo & 54.0 & 40.4 & 55.0 & 53.3 & 48.4 & 48.0 & 46.8 & 36.7 & 42.7 & 39.7 & 20.0 \\ 
        OCTOCODER & 22.0 & 20.2 & 33.0 & 28.7 & 21.8 & 16.4 & 27.2 & 25.3 & 16.0 & 29.0 & 14.0 \\ 
        OpenCodeInterpreter-DS-6.7B & 47.0 & 42.3 & 64.0 & 58.0 & 45.9 & 47.6 & 46.4 & 42.0 & 43.3 & 44.0 & 24.0 \\ 
        Phi-3-medium-4k-instruct & 40.0 & 26.9 & 48.0 & 48.7 & 30.3 & 39.2 & 31.6 & 35.3 & 33.3 & 39.0 & 13.0 \\ 
        Qwen1.5-72B-Chat & 30.0 & 29.8 & 43.0 & 44.7 & 36.3 & 30.4 & 38.0 & 39.3 & 32.7 & 40.0 & 21.0 \\ 
        WizardCoder-15B-V1.0 & 28.0 & 24.0 & 36.0 & 48.0 & 37.1 & 29.2 & 27.2 & 6.7 & 20.7 & 26.3 & 9.0 \\ 
        WizardCoder-Python-34B & 42.0 & 28.8 & 46.0 & 42.0 & 44.2 & 32.8 & 38.0 & 36.0 & 32.7 & 32.3 & 19.0 \\ 
        \ourmethod{} & 56.0 & 38.5 & 60.0 & 57.3 & 50.4 & 48.0 & 46.0 & 43.3 & 39.3 & 44.3 & 25.0 \\ 
    \bottomrule
    \end{tabular}}

\caption{Pass@1(\%) results of code generation performance across various application scenarios}
\label{tab:app_generation_result}
\end{table*}

\subsection{\ourmethod{} Result}
In \autoref{tab:extra_mcoder_result}, we show some extra \ourmethod{} Pass@1 (\%) results on multilingual code generation tasks. We evaluate the base models CodeQwen-1.5 and DeepsSeek-Coder-1.5 respectively. In addition to CodeQwen-1.5, we also selected DeepSeek-Coder-1.5-base as the base model for fine-tuning. 

\begin{table*}[ht]
\centering
\resizebox{1.0 \textwidth}{!}{
\begin{tabular}{l|ccccccccccccccccccccc}
\toprule
\textbf{Method} &  Size   & AWK  & C  & C++   & C\#   & Clisp   & Coffee  & Dart   & Elisp   & Elixir   & Erlang  & Fortran   & F\#   & Go   & Groovy   & Haskell & Html & Java  & JS & Json & Julia   \\ 
\midrule
DeepSeekCoder-1.5-base & 7B & 30.0 & 36.0 & 38.0 & 40.0 & 40.0 & 58.0 & 0.0 & 18.0 & 2.0 & 14.0 & 50.0 & 44.0 & 48.0 & 26.0 & 2.0 & 4.0 & 49.1 & 32.0 & 16.0 & 34.0 \\ 
CodeQwen-1.5 &7B & 38.0 & 40.0 & 46.0 & 42.0 & 28.0 & 56.0 & 2.0 & 14.0 & 14.0 & 0.0 & 40.0 & 46.0 & 44.0 & 32.0 & 2.0 & 0.0 & 47.2 & 52.0 & 30.0 & 60.0 \\
CodeQwen-1.5-Python & 7B & 42.0 & 48.0 & 48.0 & 12.0 & 52.0 & 68.0 & 23.5 & 22.0 & 40.0 & 62.0 & 50.0 & 42.0 & 48.0 & 68.0 & 56.0 & 32.0 & 67.9 & 52.0 & 52.0 & 54.0 \\ 
\textbf{\ourmethod{}}-DS & 7B & 34.0 & 46.0 & 50.0 & 26.0 & 30.0 & 72.0 & 19.6 & 6.0 & 26.0 & 24.0 & 58.0 & 30.0 & 48.0 & 12.0 & 26.0 & 28.0 & 67.9 & 48.0 & 62.0 & 48.0 \\ 
\textbf{\ourmethod{}}& 7B & 40.0 & 44.0 & 52.0 & 62.0 & 46.0 & 66.0 & 21.6 & 30.0 & 44.0 & 52.0 & 56.0 & 44.0 & 48.0 & 70.0 & 32.0 & 34.0 & 54.7 & 54.0 & 66.0 & 56.0 \\ 
\midrule
\textbf{Method}   & Kotlin   & Lua  & MD & Pascal  & Perl   & PHP   & Power  & Python  & R   & Racket   & Ruby   & Rust   & Scala  & Scheme   & Shell   & Swift   & Tcl   & TS  & VB   & VimL  & \textbf{Avg$_{all}$}   \\ 
\midrule
DeepSeekCoder-1.5-7B-base & 42.0 & 20.0 & 0.0 & 24.0 & 24.0 & 36.0 & 42.0 & 54.0 & 24.0 & 20.0 & 38.0 & 39.6 & 44.0 & 20.0 & 18.0 & 32.0 & 10.0 & 44.0 & 22.0 & 22.0 & 28.9 \\ 
CodeQwen-1.5 & 58.0 & 50.0 & 0.0 & 14.0 & 20.0 & 10.0 & 48.0 & 38.0 & 30.0 & 24.0 & 36.0 & 52.8 & 32.0 & 34.0 & 42.0 & 46.0 & 30.0 & 52.0 & 54.0 & 22.0 & 33.2 \\
CodeQwen-1.5-Python & 46.0 & 46.0 & 24.0 & 42.0 & 36.0 & 36.0 & 54.0 & 44.0 & 36.0 & 40.0 & 46.0 & 52.8 & 58.0 & 40.0 & 42.0 & 62.0 & 48.0 & 52.0 & 58.0 & 18.0 & 45.5 \\ 
\textbf{\ourmethod{}}-DS & 36.0 & 42.0 & 22.0 & 34.0 & 8.0 & 34.0 & 46.0 & 42.0 & 22.0 & 40.0 & 56.0 & 45.3 & 48.0 & 30.0 & 38.0 & 48.0 & 34.0 & 46.0 & 50.0 & 28.0 & 37.8 \\ 
\textbf{{\ourmethod}} & 48.0 & 52.0 & 30.0 & 42.0 & 36.0 & 32.0 & 54.0 & 44.0 & 40.0 & 36.0 & 48.0 & 52.8 & 58.0 & 44.0 & 46.0 & 64.0 & 38.0 & 52.0 & 58.0 & 20.0 & 46.7 \\ 
\bottomrule
\end{tabular}}
\caption{Additional \ourmethod{} Pass@1 (\%) results on multilingual code generation tasks. ``Avg$_{all}$" represents the average Pass@1 scores across all programming languages in the \benchmark{}. Here, \ourmethod{}-DS indicates that the fine-tuned base model is DeepSeekCoder-1.5-7b-base.}
\label{tab:extra_mcoder_result}
\vspace{-10pt}
\end{table*}

\section{Programming Grammar}

\begin{figure*}[htbp]
\begin{center}
    \includegraphics[width=1.0\textwidth]{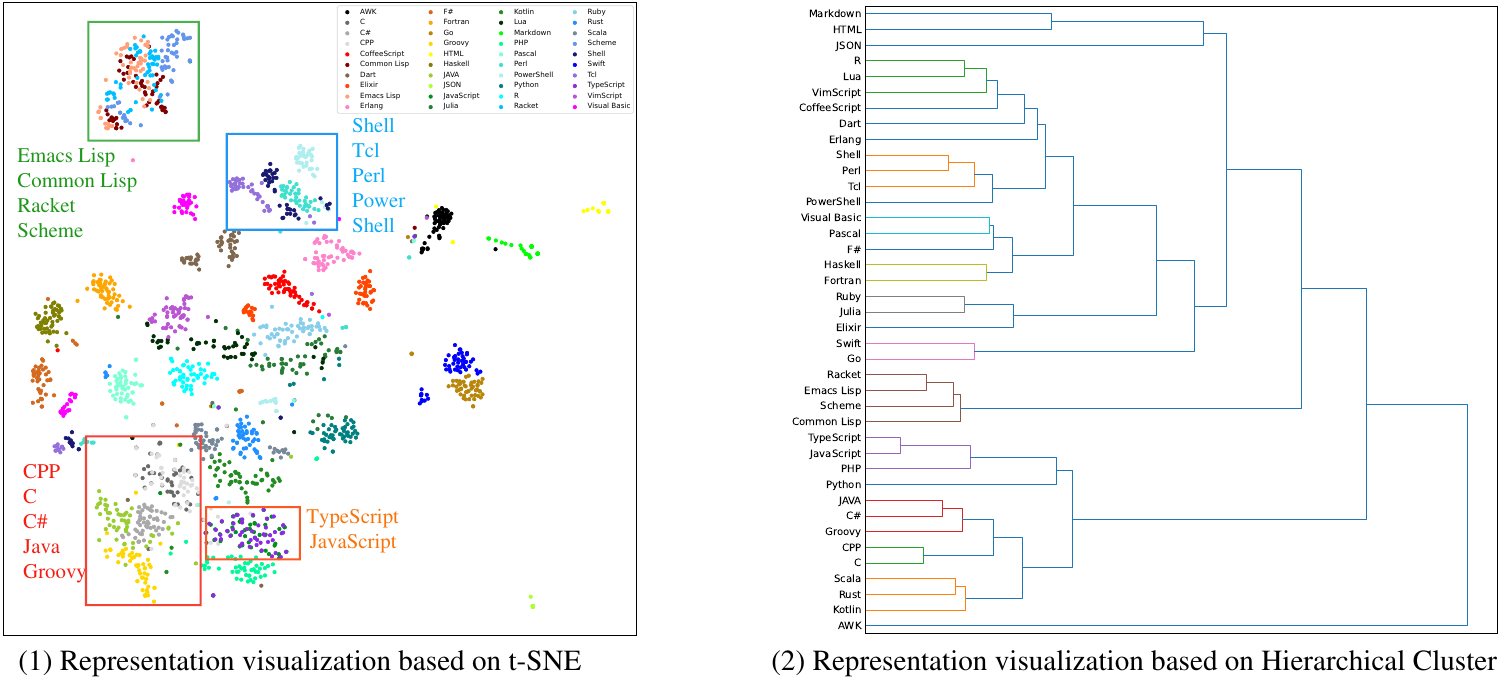}
    \caption{Analysis from the representation perspective on \benchmark{}. Languages with similar syntax have closely related representations}
    \label{fig:code_cluster}
\end{center}
\end{figure*}

As shown in \autoref{fig:code_cluster}, we analyzed the programming languages in the \benchmark{} from the representation perspective. We used CodeBERT~\citep{code_bert} to extract code representations from code snippets in \benchmark{}. These representations were visualized using t-SNE~\citep{t-sne} and hierarchical clustering~\citep{hierarchical_cluster} methods. The figure clearly shows that languages with similar syntax have closely related representations. For example, other functional programming languages similar to Common Lisp, as well as C, C++, Java, and scripting languages, exhibit high grammar similarity.

\section{Related Work}
\paragraph{Code Large Language Model.} 
In recent years, numerous large language models (LLMs) have been developed specifically for code-related tasks. For the field of soft engineering, code LLMs~\citep{code_bert,codex,bloom,AlphaCode,santacoder,incoder,codet5,opencodeinterpreter,deepseek_coder} pre-trained on billions of code snippets, such as StarCoder~\citep{starcoder, starcoder2}, CodeLlama~\citep{code_llama}, DeepSeekCoder~\citep{deepseek_coder}, and Code-Qwen~\citep{Qwen}. The development and refinement of code LLMs have been pivotal in automating software development tasks, providing code suggestions, and supporting code generation/translation. 

This segment explores the related work and key contributions in the domain of code LLMs, highlighting advancements, applications, and future directions.


To improve the performance of code generation, researchers used optimized prompts~\citep{ChatGPT_Prompt, Prompt_Programming, Instruction_Tuning, Prompting_Is_Programming}, bring test cases~\citep{Generated_Tests} and collaborative roles~\citep{Selfcollaboration}. There are also some related studies on using large language models for other code tasks, such as dynamic programming~\citep{Dynamic}, compiler optimization~\citep{Compiler}, multi-lingual prompts~\citep{CodeFuse}, and Program of Thoughts~\citep{PoT}.

\paragraph{Code Evaluation.}
In the domain of code evaluation, a rich tapestry of benchmarks~\citep{humaneval_x,codereval,arcade_nl2code,humaneval_xl,xcodeeval,BabelCode} has been woven to address the challenges of accurately assessing code quality, functionality, and efficiency, such as HumanEval~\citep{codex}, MBPP~\citep{mbpp}, their upgraded version EvalPlus~\citep{evalplus}.
Studies have explored a variety of approaches, ranging from static analysis techniques (e.g. exact match (EM) and edit similarity (ES)), which examine code without executing it, to dynamic methods that involve code execution in controlled environments (e.g. Pass@k). The current benchmarks support code models to evaluate a series of different types of tasks, such as code understanding, code repair~\citep{QuixBugs,debugbench,swe_bench,EvalGPTFix,runbugrun,he2022distribution}, code translation~\citep{CodeTransOcean}. Some recent works pay attention to the multilingual scenarios~\citep{multipl_e,odex,mbxp,codegeex,humaneval_xl,humaneval_x} by extending the existing python-only HumanEval or MBPP benchmark, such as MultiPL-E~\citep{multipl_e} and MBXP~\citep{mbxp}, which is challenged by the number of the covering languages and data leaking problem~\citep{starcoder,livecodebench}.

\end{document}